\documentclass[11pt]{article}

\newcommand{\blind}{0}

\makeatletter
\renewcommand\section{\@startsection {section}{1}{\z@}%
{-3.5ex \@plus -1ex \@minus -.2ex}%
{2.3ex \@plus.2ex}%
{\normalfont\fontfamily{phv}\fontsize{16}{19}\bfseries}}
\renewcommand\subsection{\@startsection{subsection}{2}{\z@}%
{-3.25ex\@plus -1ex \@minus -.2ex}%
{1.5ex \@plus .2ex}%
{\normalfont\fontfamily{phv}\fontsize{14}{17}\bfseries}}
\renewcommand\subsubsection{\@startsection{subsubsection}{3}{\z@}%
{-3.25ex\@plus -1ex \@minus -.2ex}%
{1.5ex \@plus .2ex}%
{\normalfont\normalsize\fontfamily{phv}\fontsize{14}{17}\selectfont}}
\makeatother
\usepackage[figuresright]{rotating}
\usepackage{graphicx}
\usepackage{amsmath}	% Advanced maths commands
\usepackage{natbib}
\bibpunct[, ]{(}{)}{;}{a}{,}{,}
\usepackage{hyperref}
\hypersetup{colorlinks=true,linkcolor=blue,citecolor=blue}
\usepackage{subfigure}
\usepackage{float}
\usepackage{amssymb}
\begin{document}

%%%%%%%%%%%%%%%%%%%%%%%%%%%%%%%%%%%%%%%%%%%%%%%%%%%%%%%%%%%%%%%%%%%%%%%%%%%%%%
\def\spacingset#1{\renewcommand{\baselinestretch}%
{#1}\small\normalsize} \spacingset{1}
%%%%%%%%%%%%%%%%%%%%%%%%%%%%%%%%%%%%%%%%%%%%%%%%%%%%%%%%%%%%%%%%%%%%%%%%%%%%%%

\if0\blind
{\title{Effect of dark matter halos on the orbital and escape dynamics of barred galaxies}
\author{Debasish Mondal$^{a,*}$ and Tanuka Chattopadhyay$^{a,\dagger}$\\
$^a$Department of Applied Mathematics, University of Calcutta,\\ $92$ A. P. C. Road, Kolkata $700009$, India\\
E-mail: $^*$dmappmath$\_$rs@caluniv.ac.in (Corresponding Author),\\
$^\dagger$tchatappmath@caluniv.ac.in}
\date{}
\maketitle
}\fi
\bigskip

\begin{abstract}
This study examines the effect of dark matter halos on the orbital and escape dynamics of stars in the central region of barred galaxies. A three-dimensional gravitational model with a central bulge, bar, disc, and dark matter halo (or simply dark halo) has been set up and analyzed from the viewpoint of escape in open Hamiltonian systems for this purpose. Additionally, this model has been examined separately for the dark halo profiles: oblate and NFW. In both circumstances, an escape mechanism has been identified near the saddle points of the phase space, which correspond to the bar ends. The escaping motion of stars is seen using orbital maps and Poincaré surface section maps generated in various phase planes. Finally, the relationship between chaos and dark halo parameters such as mass, size, circular velocity, and nature has been studied. Our findings suggest that oblate dark halos are preferred over NFW dark halos for justifying the formation of full-fledged spiral arms and extended distribution of dark halos in giant spiral galaxies with supermassive black holes (SMBHs) at their centers. Again, the oblate dark halos well justify the emergence of less prominent or poor spiral arms and the core-dominated distribution of dark halos in dwarf and LSB galaxies in the absence of central SMBHs. On the other hand, extreme central baryonic feedback is required for the NFW halos to generate spiral patterns, and such dark halos should be preferred for galaxies with extremely energetic centers.
\end{abstract}

\noindent%
{\it Keywords:} Galaxy: kinematics and dynamics;  Galaxies: bar;  Galaxies: halos;  Chaos.

%\newpage
\spacingset{1.5} % DON'T change the spacing!

\section{Introduction}
\label{sec:1}
The distribution of dark matter particles around field galaxies, galaxy groups, and galaxy clusters is known as dark matter halos (or simply dark halos). These non-baryonic (mostly) dark matter particles are one of the most important aspects of the universe's cosmic evolution. This is one of the most essential parameters for studying galaxy formation and evolution. According to the findings of the Planck collaboration \citep{Planck2016}, dark matter particles account for approximately $\frac{5}{6}\text{th}$ of the universe's mass. The inconsistencies between the anticipated mass and measured luminous mass of numerous galaxies bring them into the picture for the first time \citep{Kapteyn1922, Freeman1970, Rubin1980, Zwicky2009}. A vast dark halo is embedded over the visible components of giant spiral galaxies like the Milky Way, M31, NGC 1365, and others, and its density steadily decreases as one moves away from the centre. The radial span of such dark halos is about 100 to 200 kpc, whereas the luminous portion of the galaxy has a radius of only 30 to 35 kpc \citep{Klypin2002, Bhattacharjee2014, Huang2016}. Again, for the dwarf and Low-Surface-Brightness (LSB) galaxies, the structures of dark halos are mostly core-dominated \citep{Blok2002, Simon2005, Swaters2011}. There are a few galaxies with nearly no dark matter, such as NGC 1052-DF2 and NGC 1052-DF4 \citep{Haslbauer2019, Yang2020}. Hence, the distribution of dark halos varies across galaxy morphologies and has an immense influence on the kinematic and structural properties of galaxies \citep{Weinberg2002, Salucci2019, Thob2019}.

\indent Dark halos are closely associated with the formation of stars and galaxies in the early universe. In view of the cosmological Lambda Cold Dark Matter ($\Lambda$CDM) model, dark matters are collision-less and evolve under the influence of gravity. Initial small-scale density perturbations in the universe grow linearly and, upon reaching a threshold value, expansion halts and further collapses to form gravitationally bound structures in the form of small dark halos. Further, these small dark halos were merged to form a single virialized dark halo structure, and that reveals some structure in the form of sub dark halos. These sub halos further gravitationally interact with baryonic matter in order to overcome the thermal energy of baryonic interactions. This allowed baryonic matter to collapse into nonlinear gravitationally bound structures in the form of the first stars and galaxies in the universe \citep{Blumenthal1984, White1991, Springel2005, Popesso2015, Wechsler2018}. 

\indent Cosmological \textit{N}-body simulations of collision-less dark matter particles are widely used to model the structure of dark halos \citep{Moore1998, Yoshida2000, Hahn2013, Fischer2021}. These \textit{N}-body simulations over predict the number of sub-halos in the subsequent cosmic environment, known as the `missing satellites or dwarf galaxy problem' \citep{Kazantzidis2004, Tanaka2018}. Moreover, cosmological \textit{N}-body simulations predict that dark halos have a structure akin to the Navarro–Frenk–White (NFW) profile \citep{Navarro1996}. The central density of the NFW profile is divergent (infinite) and has a cuspy distribution near the center. Although this NFW profile agrees well with the dark halo profiles of giant galaxy clusters, it is proven to be unobserved across a variety of galaxy morphologies. In contrast to the results produced from cosmological \textit{N}-body simulations, rotation curves of the most dwarf and LSB galaxies reveal flat, i.e., core-dominated dark halos. The later Einasto profile \citep{Einasto1965} provides an alternative to the NFW profile with finite central density, but it is still cuspier than the dwarf and LSB galaxies' dark halos. This discrepancy between the predicted dark halo profiles of \textit{N}-body simulations and the observed dark halo profiles of dwarf and LSB galaxies is still a matter of debate and is known as the `cuspy halo problem' or `core-cusp problem \citep{Jing2000, Blok2010, Ogiya2011, Popolo2021}. To shed light on this topic here, we studied a barred galaxy model under two different dark halo profiles, viz., oblate and NFW.

\indent In \citet{Mondal2021}, a barred galaxy model has been analyzed separately for strong and weak bars to study the orbital and escape dynamics of stars inside the central barred region and its subsequent effects in terms of the formation of spiral arms or inner disc rings. In the present work, we analyzed the role of dark halos on the orbital and escape dynamics inside the central barred region. Also, we discussed the possibilities of subsequent structure formations based upon the strength of the dark halo component. This study has been done from the viewpoint of escape in an open Hamiltonian system \citep{Aguirre2001}, where unstable manifolds in the phase space are associated with stellar escapes, and such escape mechanisms are visualized via Poincaré surface section maps \citep{Birkhoff1927} in different two-dimensional phase planes. Also, the chaotic nature of the stellar orbits is measured through the Maximal Lyapunov Exponent (MLE) values \citep{Strogatz1994}.

\indent The majority of early studies of barred galaxies' orbital and escape dynamics were based on bar strength and the formation of spiral arms or inner disc rings as a result of bar-driven escape. These studies also look into the role of normally hyperbolic invariant manifolds in these stellar escape mechanisms \citep{Navarro2001, Romero2006, Voglis2006, Romero2007, Zotos2011, Jung2016, Sanchez2016, Efthymiopoulos2019}. The impact of dark halos on escape and chaotic dynamics within the barred region has received little attention so far. Our research indoctrinates about the impact of dark halos on orbital and escape dynamics inside the central barred region, as well as their impact on galaxies' structure formation. This also provides a fair glimpse of the 'cuspy halo problem'.

\indent In this study, we employ a three-dimensional gravitational model of barred galaxies . This model is made up of four components: a spherical bulge, a strong bar, a flat disc, and a dark halo. For a comparison analysis, two dark halo profiles were chosen: (i) an oblate profile or flat dark halo \citep{Binney2008, Zotos2012, Mondal2021}, and (ii) the NFW profile or cuspy dark halo \citep{Navarro1996}. Orbital maps are drawn for suitable initial conditions in both dark halo models, and corresponding MLE values are derived to measure orbital chaos. For different escape energy values, Poincaré surface section maps in the $x - y$ and $x - p_x$ phase planes are plotted to study the escaping trends inside the barred region. Finally, the effect of different dark halo parameters, such as mass, size, circular velocity, and nature, on orbital chaos is investigated.

\indent There are four sections to the overall work. The introduction is described in Section \ref{sec:1}. The mathematical elements of the gravitational model are described in Section \ref{sec:2}, which is divided into three sub-sections. The model for an oblate dark halo is described in Section \ref{sec:2.1}, the model for the NFW dark halo is described in Section \ref{sec:2.2}, and thorough comparisons between the two dark halo models are given in Section \ref{sec:2.3}. Section \ref{sec:3} contains three sub-sections devoted to numerical research. Section \ref{sec:3.1} includes orbital maps in the $x - y$ plane, Section \ref{sec:3.2} includes Poincaré surface section maps in the $x - y$ and $x - p_x$ planes, and Section \ref{sec:3.3} includes MLE variation with various dark halo parameters. Section \ref{sec:4} contains the final discussion and conclusions.

\section{Gravitational model}
\label{sec:2}
We consider a three-dimensional gravitational model of barred galaxies. Our model components are the central bulge, bar, disc and extended dark halo. We investigate the role of dark halos on the orbital and escape dynamics of stars inside the central barred region. Here, all the modelling and numerical calculations are done in a Cartesian coordinate system. Let $\Phi_\text{t}(x,y,z)$ be the total galactic potential and $\rho_\text{t}(x,y,z)$ be the associated density. This potential-density pair is related by the Poisson equation, 
$$\nabla^2 \Phi_\text{t}(x,y,z) = 4 \pi G \rho_\text{t}(x,y,z).$$ 
Now, $$\Phi_\text{t}(x,y,z) = \Phi_\text{B}(x,y,z) + \Phi_\text{b}(x,y,z) + \Phi_\text{d}(x,y,z) + \Phi_\text{h}(x,y,z),$$ where $G$ is the gravitational constant and the potentials for the bulge, bar, disc and dark halo are $\Phi_\text{B}$, $\Phi_\text{b}$, $\Phi_\text{d}$, $\Phi_\text{h}$, respectively. Let $\vec{{\Omega}}_\text{b} \equiv (0,0,\Omega_\text{b})$ be the constant rotational velocity (in clockwise sense along $z$ - axis) of the bar. In this rotational reference frame of the bar, the effective potential is,
\begin{equation}
\label{eq:1}
\Phi_\text{eff}(x,y,z) = \Phi_\text{t}(x,y,z) - \frac{1}{2} \Omega_\text{b}^2 (x^2 + y^2).
\end{equation}

\noindent For a test particle (star) of unit mass, the Hamiltonian ($H$) of the given conservative system is,
\begin{equation}
\label{eq:2}
H = \frac{1}{2} (p_x^2 + p_y^2 + p_z^2) \; + \; \Phi_\text{t}(x,y,z) \; - \; \Omega_\text{b} L_z = E,
\end{equation}
where $\vec{r} \equiv (x,y,z)$, $\vec{p} \equiv (p_x,p_y,p_z)$ and $\vec{L} \; (=\vec{r} \times \vec{p}) \equiv (0,0,L_z = x p_y - y p_x)$ are the position, linear momentum vector and angular momentum vector of test particle at time $t$, respectively. This is a conservative system and for that $H = E$ (system's total energy). So, system's governing equations, i.e., Hamilton's equations of motion are, 

\begin{equation}
\label{eq:3}
\begin{split}
\dot{x} = p_x + \Omega_\text{b} y,\\
\dot{y} = p_y - \Omega_\text{b} x,\\
\dot{z} = p_z,\\
\dot{p_x} = - \frac{\partial \Phi_\text{t}}{\partial x} + \Omega_\text{b} p_y,\\
\dot{p_y} = - \frac{\partial \Phi_\text{t}}{\partial y} - \Omega_\text{b} p_x,\\
\dot{p_z} = - \frac{\partial \Phi_\text{t}}{\partial z},\\
\end{split}
\end{equation}
where $`\cdot$' represents the time derivative $\frac{\mathrm{d}}{\mathrm{dt}}$. Now, the Lagrangian (or equilibrium) points of this Hamiltonian system are solutions of the following equations,
\begin{equation}
\label{eq:4}
\frac{\partial \Phi_\text{eff}}{\partial x} = 0, \;\; 
\frac{\partial \Phi_\text{eff}}{\partial y} = 0, \;\; 
\frac{\partial \Phi_\text{eff}}{\partial z} = 0.
\end{equation}

\subsection{Model 1}
\label{sec:2.1} 
The potential forms of the bulge, bar, disc, and dark halo are as follows: 
\begin{itemize}
\item Bulge: In this model, we consider a massive dense bulge rather than a central Super Massive Black Hole (SMBH) in order to exclude all relativistic effects. For this dense spherical bulge, we use Plummer potential \citep{Plummer1911},
$$\Phi_\text{B}(x,y,z) = - \frac{G M_\text{B}}{\sqrt{x^2 + y^2 + z^2 + c_\text{B}^2}},$$
where $M_\text{B}$ is the bulge mass and $c_\text{B}$ is the scale length.

\item Bar: For the central stellar bar, we consider an anharmonic mass-model potential, which resembles a strong bar \citep{Mondal2021}. Form of this potential is,
$$\Phi_\text{b}(x,y,z) = - \frac{G M_\text{b}}{\sqrt{x^2 + \alpha^2 y^2 + z^2 + c_\text{b}^2}},$$ 
where $M_\text{b}$ is the bar mass, $\alpha$ is the flattening parameter and $c_\text{b}$ is the scale length.

\item Disc: For the flattened disc, we use Miyamoto and Nagai potential \citep{Miyamoto1975},
$$\Phi_\text{d}(x,y,z) = - \frac{G M_\text{d}}{\sqrt{x^2 + y^2 + (k + \sqrt{h^2 + z^2})^2}},$$
where $M_\text{d}$ is the disc mass and $k$, $h$ are the corresponding horizontal and vertical scale lengths, respectively.

\item Dark halo: For the extended dark halo, we consider a flat dark halo profile (see Fig. \ref{fig:2a}). For this, we use an oblate dark halo potential \citep{Binney2008, Zotos2012, Mondal2021},
$$\Phi_\text{h}(x,y,z) = \frac{v_0^2}{2} \; \ln(x^2 + \beta^2 y^2 + z^2 + c_\text{h}^2),$$
where $v_0$ is the circular velocity of the dark halo, $\beta$ is the flattening parameter and $c_\text{h}$ is the scale length.
\end{itemize}

In this gravitational model without loss of any generality, we consider $G$ = 1 and adopt the following system of units -- unit of length: $1$ kpc, unit of mass: $2.325 \times 10^7 M_\odot$, unit of time: $0.9778 \times 10^8$ yr, unit of velocity: $10$ km $\text{s}^{-1}$, unit of angular momentum per unit mass: $10$ km $\text{s}^{-1}$ $\text{kpc}^{-1}$, unit of energy per unit mass: $100$ $\text{km}^2$ $\text{s}^{-2}$ \citep{Jung2016}. Table \ref{tab:1} shows the values of all physical parameters of model $1$ \citep{Zotos2012, Jung2016} according to these scaling relations.
\begin{table}[H]
\centering	
\begin{tabular}{|c|c||c|c|}
\hline
Parameter          & Value & Parameter    & Value\\
\hline
\hline
$M_\text{B}$       & 400   & $M_\text{d}$ & 7000  \\
$c_\text{B}$       & 0.25  & $k$          & 3\\
$M_\text{b}$       & 3500  & $h$          & 0.175\\
$\alpha$           & 2     & $v_0$        & 15\\
$c_\text{b}$       & 1     & $\beta$      & 1.3\\
$\Omega_\text{b}$  & 1.25  & $c_\text{h}$ & 20\\
\hline
\end{tabular}
\caption{Model $1$: Physical parameter values.}
\label{tab:1}
\end{table}

\noindent This model has five Lagrangian points: $L_1$, $L_2$, $L_3$, $L_4$ and $L_5$. Section 2.1 of \citet{Mondal2021} contains the locations of the Lagrangian points, their types, and the corresponding energy values.

\subsection{Model 2}
\label{sec:2.2}
\begin{figure}[H]
\centering
\includegraphics[width=0.5\columnwidth]{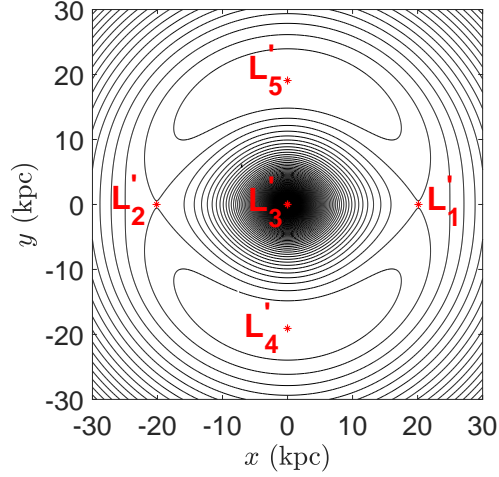}
\caption{Model $2$: The isoline contours of $\Phi_\text{eff}(x,y,z)$ in the $x - y$ plane for $z = 0$, where the locations of the Lagrangian points are marked in red. For $E < E_{L_1^{'}} (= -1230.63004977)$ orbits follow bounded motions around $L_3^{'}$, and for $E > E_{L_1^{'}}$ orbital escape is possible through the exit channels near ${L_1^{'}}$ and ${L_2^{'}}$.}
\label{fig:1}
\end{figure}

In this part, we use the same three-dimensional gravitational model as discussed in Section \ref{sec:2.1}. Here we change only the form of dark halo potential by considering a cuspy dark halo profile (see Fig. \ref{fig:2a}). For this, we choose the NFW potential \citep{Navarro1996}, 
$$\Phi_\text{h}(x,y,z) = - \frac{G M_\text{vir}}{\ln (1 + c) - \frac{c}{1 + c}} \frac{\ln (1 + \frac{r}{c_\text{h}})}{r},$$ 
where $r^2 = x^2 + y^2 + z^2$, $M_\text{vir}$ is the virial mass of the dark halo, $c_\text{h}$ is the scale length and $c$ is the concentration parameter. For this dark halo potential, we use $M_\text{vir} = 20000$ \citep{Jung2016}, $c = 15$ and values of other model parameters remain the same as given in Table \ref{tab:1}. The locations and types of the Lagrangian points for this model, namely $L_1^{'}$, $L_2^{'}$, $L_3^{'}$, $L_4^{'}$ and $L_5^{'}$ are given in Fig. \ref{fig:1} and Table \ref{tab:2}.

\begin{table}[H]	
\centering
\begin{tabular}{|c|c|c|}
\hline
Lagrangian Point & Location      & Type\\
\hline
\hline
$L_1^{'}$ & $(20.13638156,0,0)$  & Index-1 Saddle\\
$L_2^{'}$ & $(-20.13638156,0,0)$ & Index-1 Saddle\\
$L_3^{'}$ & $(0,0,0)$            & center\\
$L_4^{'}$ & $(0,-19.07590847,0)$ & Index-2 Saddle\\
$L_5^{'}$ & $(0,19.07590847,0)$  & Index-2 Saddle\\
\hline
\end{tabular}
\caption{Model $2$: Lagrangian point locations and types.}
\label{tab:2}
\end{table}

The values of Jacobi integral of motion at $L_1^{'}$ and $L_2^{'}$ are identical and that value is $E_{L_1^{'}} = -1230.63004977 = E_{L_2^{'}}$. Similarly, the values of Jacobi integral of motion at $L_4^{'}$ and $L_5^{'}$ are identical and that value is $E_{L_4^{'}} = -1141.60154080 = E_{L_5^{'}}$. Also, the Jacobi integral of motion value at $L_3^{'}$ is $E_{L_3^{'}} = - \infty$. The nature of orbits in different energy domains are,
\begin{enumerate}	
\item $E_{L_3^{'}} \leq E < E_{L_1^{'}}$: In this energy domain, stellar orbits follow bounded motions inside the barred region.
\item $E = E_{L_1^{'}}$: This is the threshold energy for escape through the bar ends.
\item $E > E_{L_1^{'}}$: In this energy domain, orbital escape of stars is possible through the two symmetrical escape channels that exist near ${L_1^{'}}$ and ${L_2^{'}}$ depending upon the initial starting point.
\end{enumerate}

\subsection{Comparison between model 1 and model 2}  
\label{sec:2.3}
\begin{figure}[H]
\centering
\subfigure[Evolution of the dark halo density ($\rho_\text{h}$) with radius ($R = \sqrt{x^2 + y^2}$) for $z = 0$]{\label{fig:2a}\includegraphics[width=0.49\columnwidth]{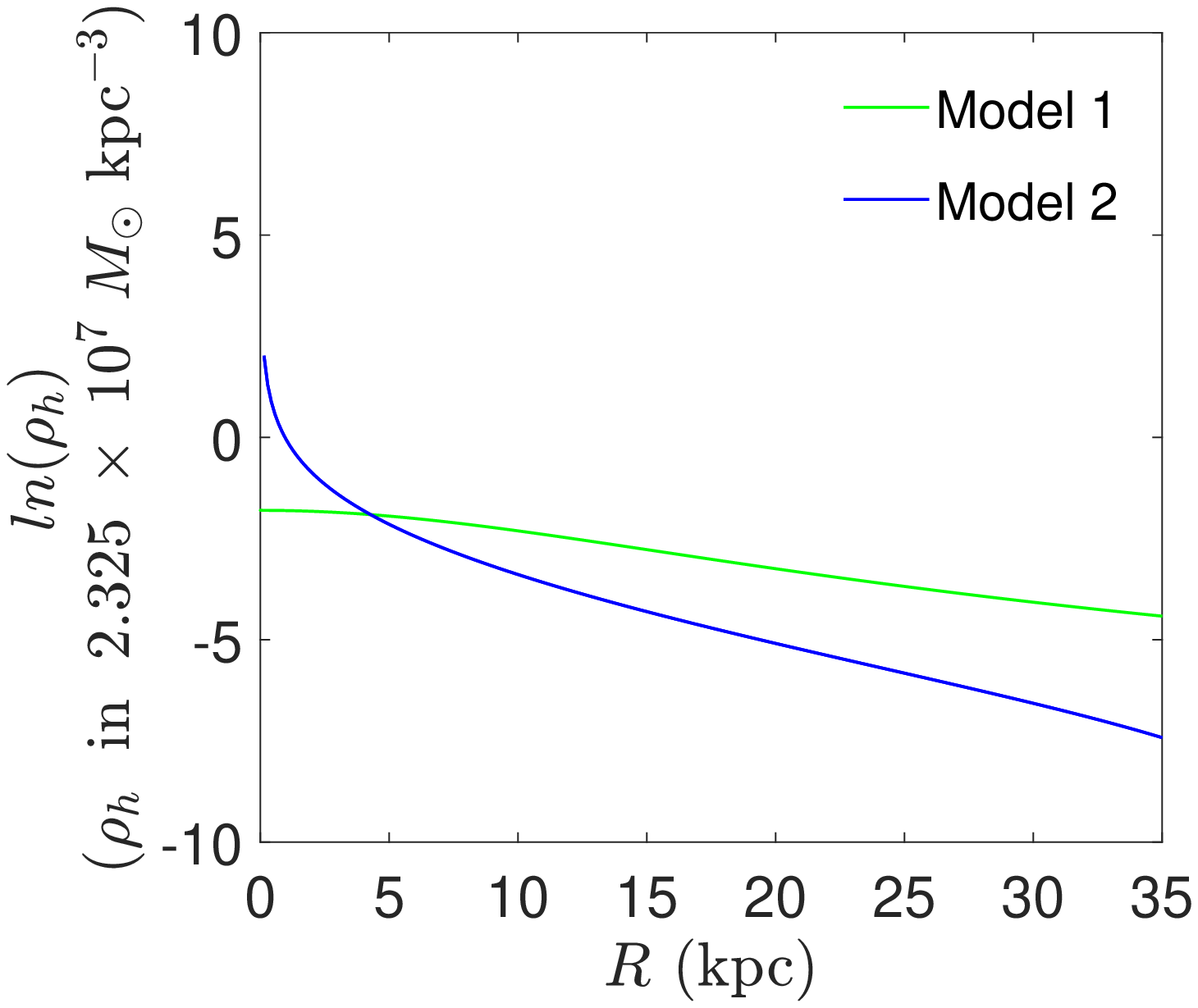}}
\subfigure[Rotation curve]{\label{fig:2b}\includegraphics[width=0.49\columnwidth]{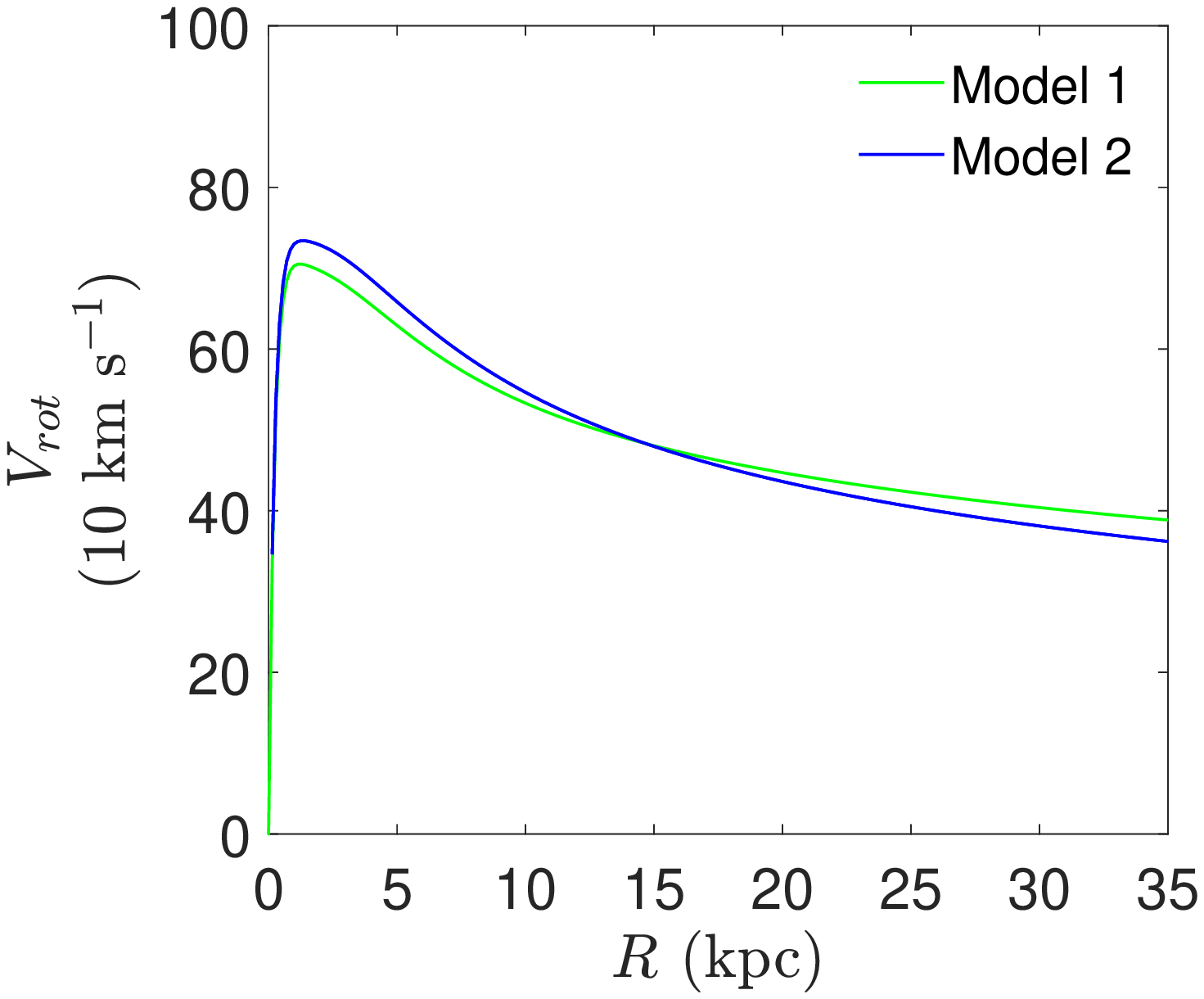}}
\subfigure[Evolution of the radial force component ($F_R$) with radius ($R = \sqrt{x^2 + y^2}$) for $z = 0$]{\label{fig:2c}\includegraphics[width=0.49\columnwidth]{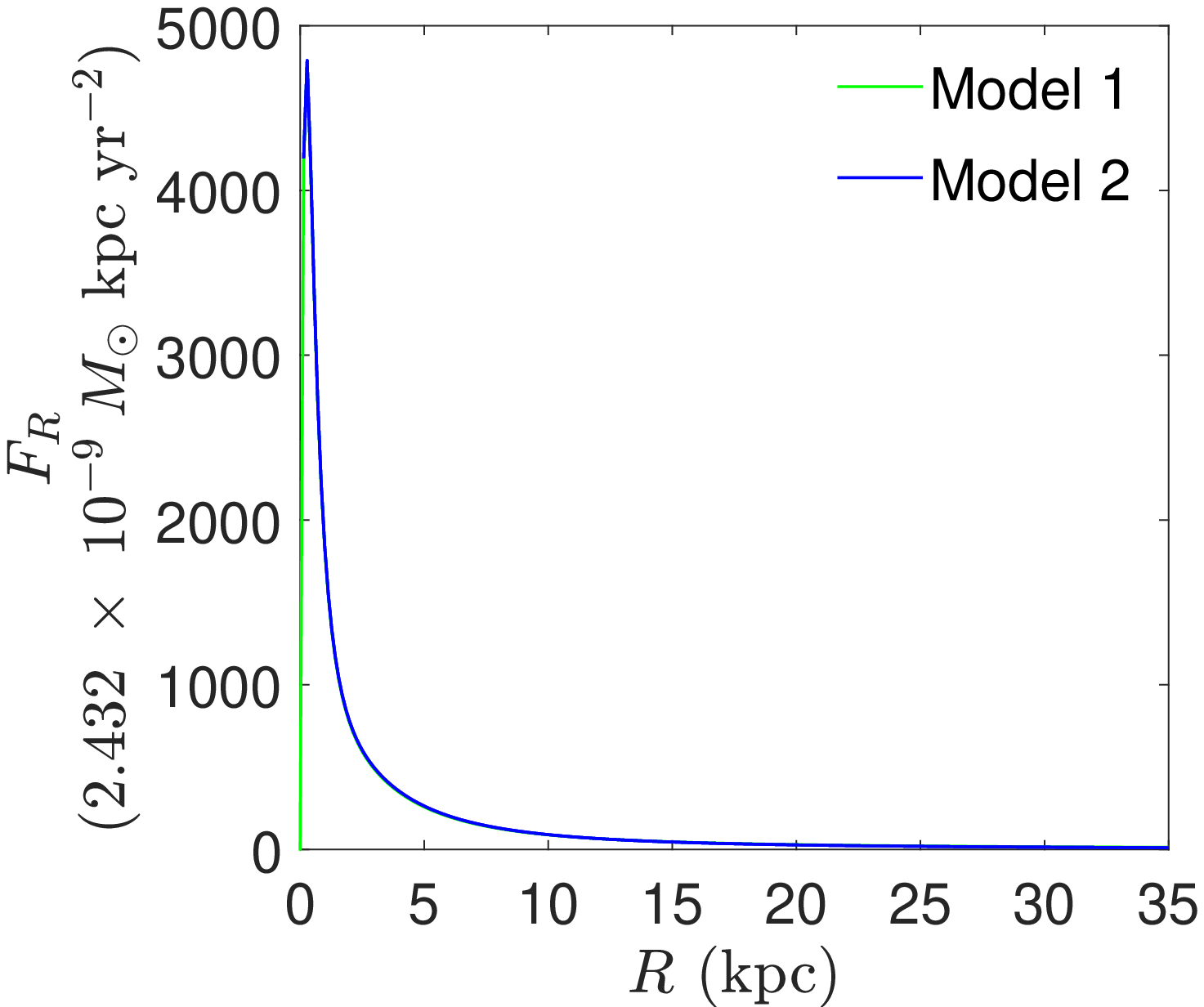}}
\subfigure[Evolution of the tangential force component ($F_\theta$) with radius ($R = \sqrt{x^2 + y^2}$) for $z = 0$]{\label{fig:2d}\includegraphics[width=0.49\columnwidth]{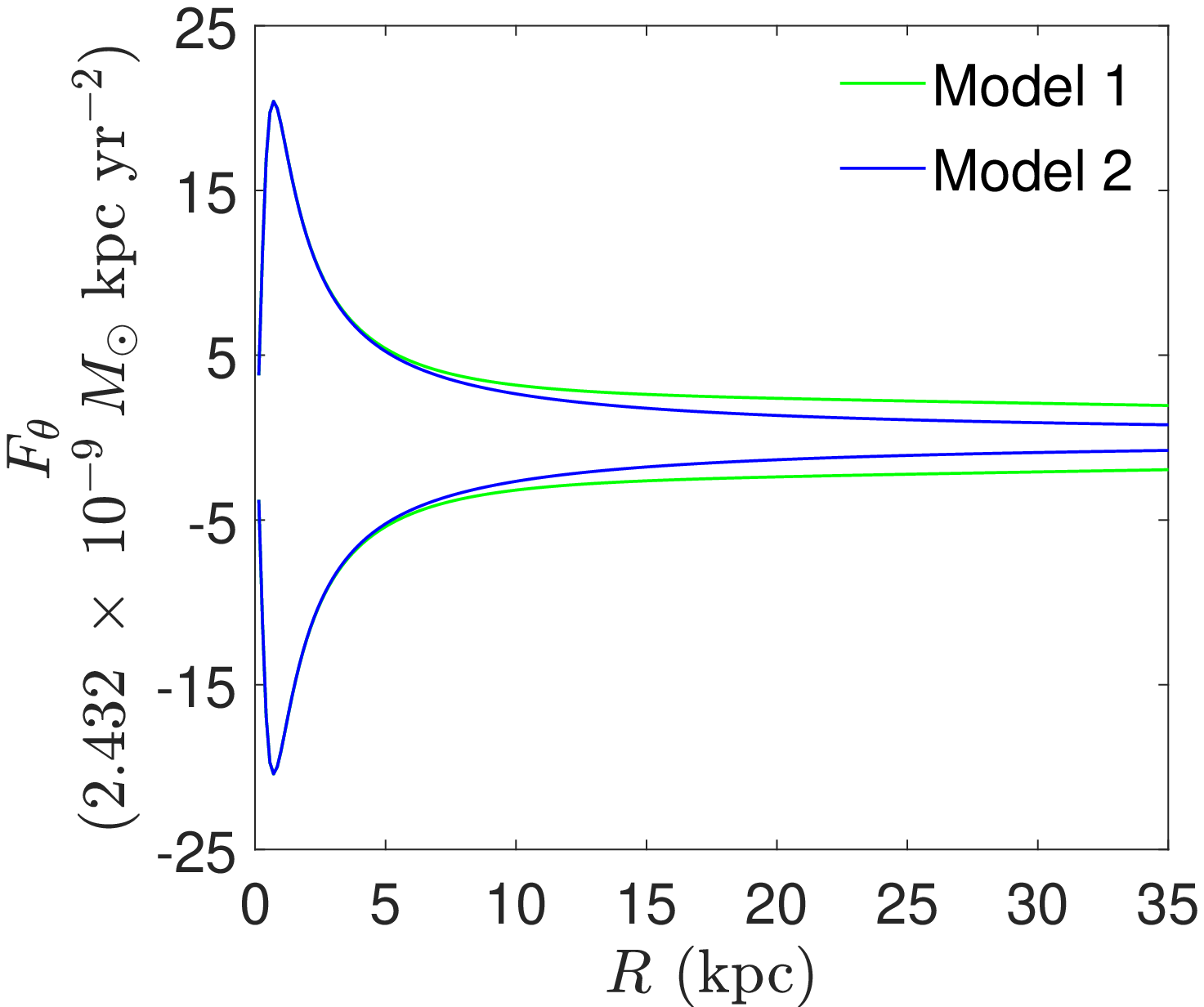}}
\caption{Model $1$ versus model $2$.}
\label{fig:2}
\end{figure}

The characteristic comparisons between model $1$ and model $2$ are as follows: 

\begin{itemize}
\item Dark halo density: For both the models, the evolution of the dark halo density ($\rho_\text{h}$) with radius ($R = \sqrt{x^2 + y^2}$) for $z = 0$ is shown in Fig. \ref{fig:2a}. In this figure, we observed that the dark halo density of model $1$ resembles a flat distribution of matter, while the dark halo density of model $2$ resembles a cuspy distribution of matter.

\item Rotation curve: For both models, the evolution of the circular velocity ($V_{\text{rot}} = \sqrt{R \frac{\partial \Phi_\text{t}}{\partial R}}$) with radius ($R = \sqrt{x^2 + y^2})$ for $z = 0$ is shown in Fig. \ref{fig:2b}. In this figure, we observed that within the barred region, the rotational velocity of model $2$ is slightly ahead of model $1$ but outside the barred region, a reverse trend is observed.

\item Radial force: For both models, the evolution of the radial force ($F_R = \frac{\partial \Phi_\text{t}}{\partial R}$) with radius ($R = \sqrt{x^2 + y^2}$) for $z = 0$ is shown in Fig. \ref{fig:2c}. In this figure, we observed that the radial force components of both models are almost identical. Also, distributions become steeper near the galactic center and gradually approach the horizontal axis for larger values of $R$.  

\item Tangential force: For both models, the evolution of the tangential force ($F_\theta = \frac{1}{R} \frac{\partial \Phi_\text{t}}{\partial \theta}$) with radius ($R = \sqrt{x^2 + y^2}$) for $z = 0$ is shown in Fig. \ref{fig:2d}. Where $\theta$ denotes the amplitude of $(x,y)$. In this figure, we observed that the tangential force components of both models are identical within the central bulge, but beyond that, the tangential force component of model $1$ slightly dominates over the tangential force component of model $2$.
\end{itemize}	

\section{Numerical studies}
\label{sec:3}
To study the orbital and escape dynamics of stars inside the central barred region, i.e., along the galactic plane of the bar, we set $z = 0 = p_z$. In the energy domain $E \geq E_{L_1}$ (or $E \geq E_{L_1^{'}}$), stellar orbits may be able to escape along the symmetrical escape channels that exist near the bar ends, i.e., $L_1$, $L_2$ (or $L_1^{'}$, $L_2^{'}$) (see Fig. 1 of \citet{Mondal2021} and Fig. \ref{fig:1}). Due to symmetry in the effective potential, it is sufficient to study only near $L_1$ (or $L_1^{'}$) to track the escape dynamics of stars. For simpler numerical calculations, we use the dimensionless energy parameter $C$ \citep{Mondal2021} given as,
\begin{equation*}
C = \frac{E_{L_1} (\text{or \;} E_{L_1^{'}}) - E}{E_{L_1} (\text{or \;} E_{L_1^{'}})} = \frac{E_{L_2} (\text{or \;} E_{L_2^{'}}) - E}{E_{L_2} (\text{or \;} E_{L_2^{'}})}.
\end{equation*} 
$$(\because E_{L_1} = E_{L_2} \; \text{and} \; E_{L_1^{'}} = E_{L_2^{'}})$$

\indent Now $C = 0$ is the threshold energy for escape from the barred region. So, for $C > 0$ orbits may escape through the symmetrical escape channels depending upon their regular or chaotic character. To study this escaping motion, our tested energy levels are $C = 0.01$ and $C = 0.1$. Also, all the chosen initial conditions are restricted within the Lagrange radius. This restricted region for model $1$ is: $x_{0}^2 + y_{0}^2 \leq r_{L_1}^2$, where $r_{L_1}$ is the radial length of $L_1$ and $(x_0, y_0)$ is an initial condition in the $x - y$ plane. Similarly for model $2$, that restricted region is: $x_{0}^2 + y_{0}^2 \leq r_{L_1^{'}}^2$, where $r_{L_1^{'}}$ is the radial length of $L_1^{'}$.  

\indent The chaotic nature of stellar orbits is evaluated with the help of the chaos detector MLE \citep{Mondal2021}. To numerically solve the system of differential Eqs. (\ref{eq:3}), we use a set of $\tt{MATLAB}$ scripts, where we use the $\tt{ode45}$ $\tt{MATLAB}$ package with a small time step ($\Delta t$) $= 10^{-2}$. The age of stellar bars is typically $10^{10}$ years \citep{Sharma2019}, which is nearly equivalent to $10^2$ time units (our total numerical integration time). All the calculations presented here are corrected up to eight decimal places, and all the graphics are produced in the $\tt{MATLAB-2015a}$ environment.

\subsection{Orbital structures}
\label{sec:3.1}
\begin{figure}[H]
\centering
\subfigure[Non-escaping chaotic orbit for $C = 0.01$ with $(x_0,y_0,p_{x_0})$ $\equiv (5,0,15)$]{\label{fig:3a}\includegraphics[width=0.49\columnwidth]{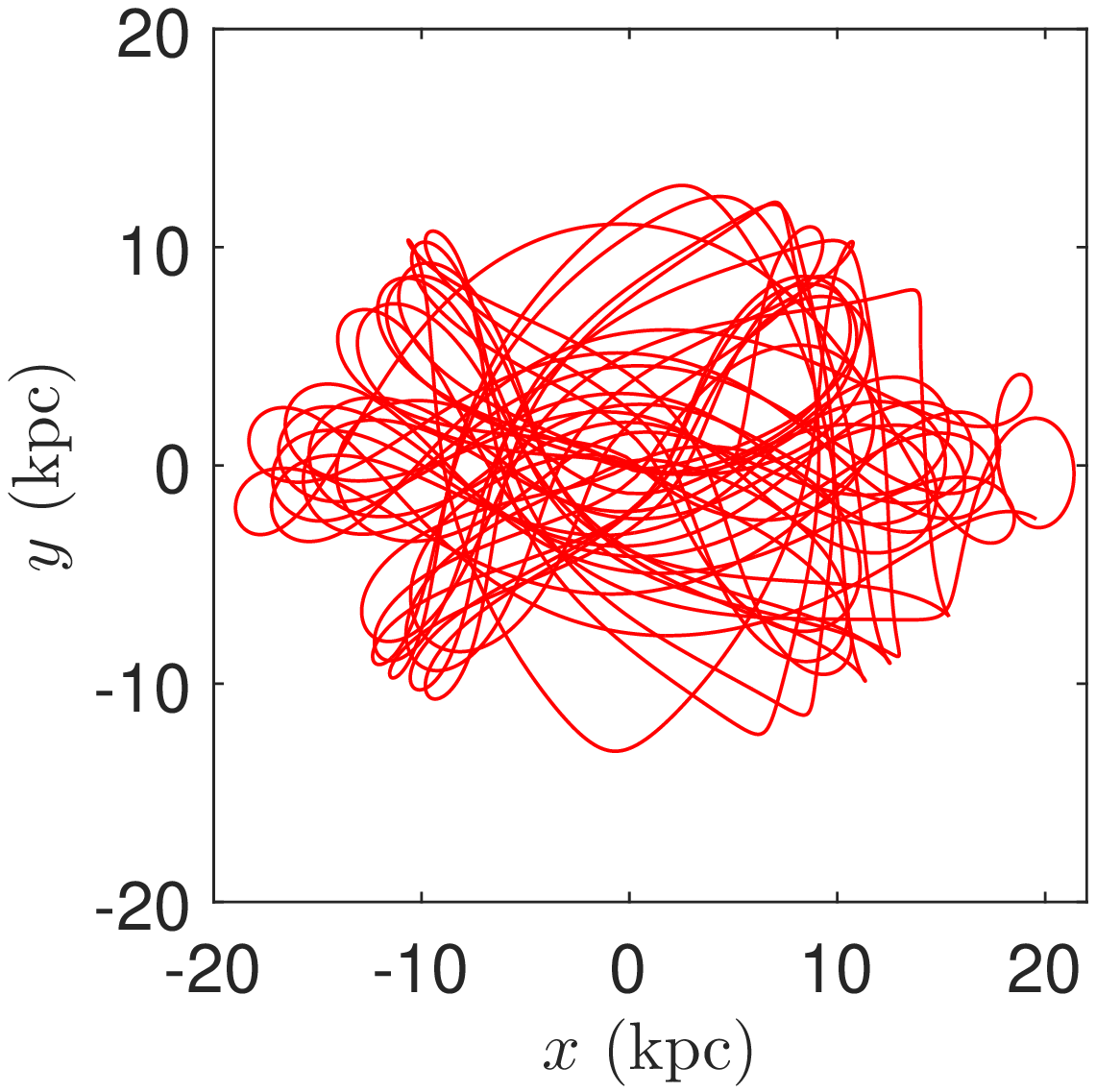}}
\subfigure[Non-escaping chaotic orbit for $C = 0.1$ with $(x_0,y_0,p_{x_0})$ $\equiv (5,0,15)$]{\label{fig:3b}\includegraphics[width=0.49\columnwidth]{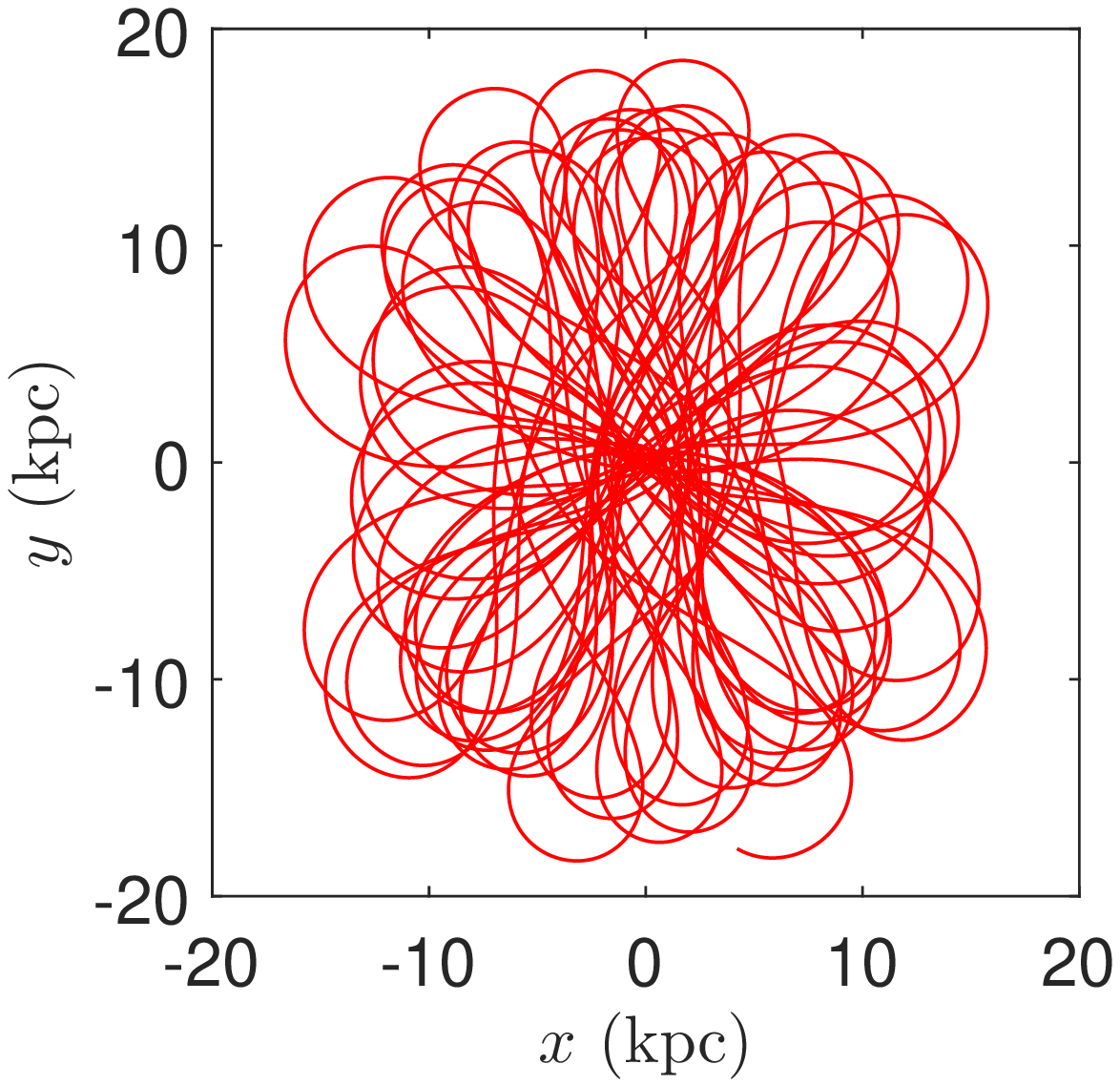}}
\subfigure[Non-escaping retrograde quasi-periodic rosette orbit for $C = 0.01$ with $(x_0,y_0,p_{x_0})$ $\equiv (-5,0,15)$]{\label{fig:3c}\includegraphics[width=0.49\columnwidth]{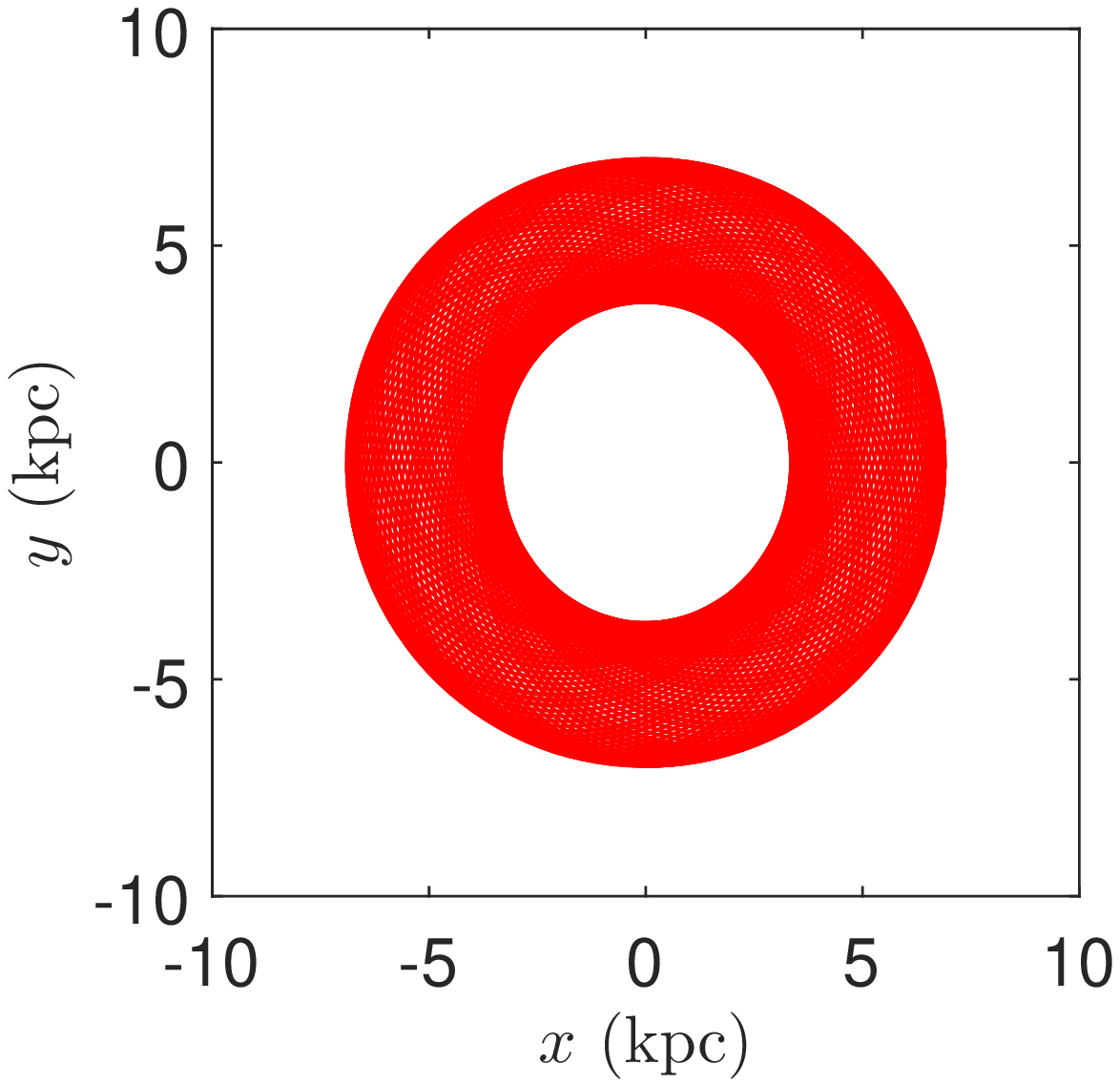}}
\subfigure[Non-escaping retrograde quasi-periodic rosette orbit for $C = 0.01$ with $(x_0,y_0,p_{x_0})$ $\equiv (-5,0,15)$]{\label{fig:3d}\includegraphics[width=0.49\columnwidth]{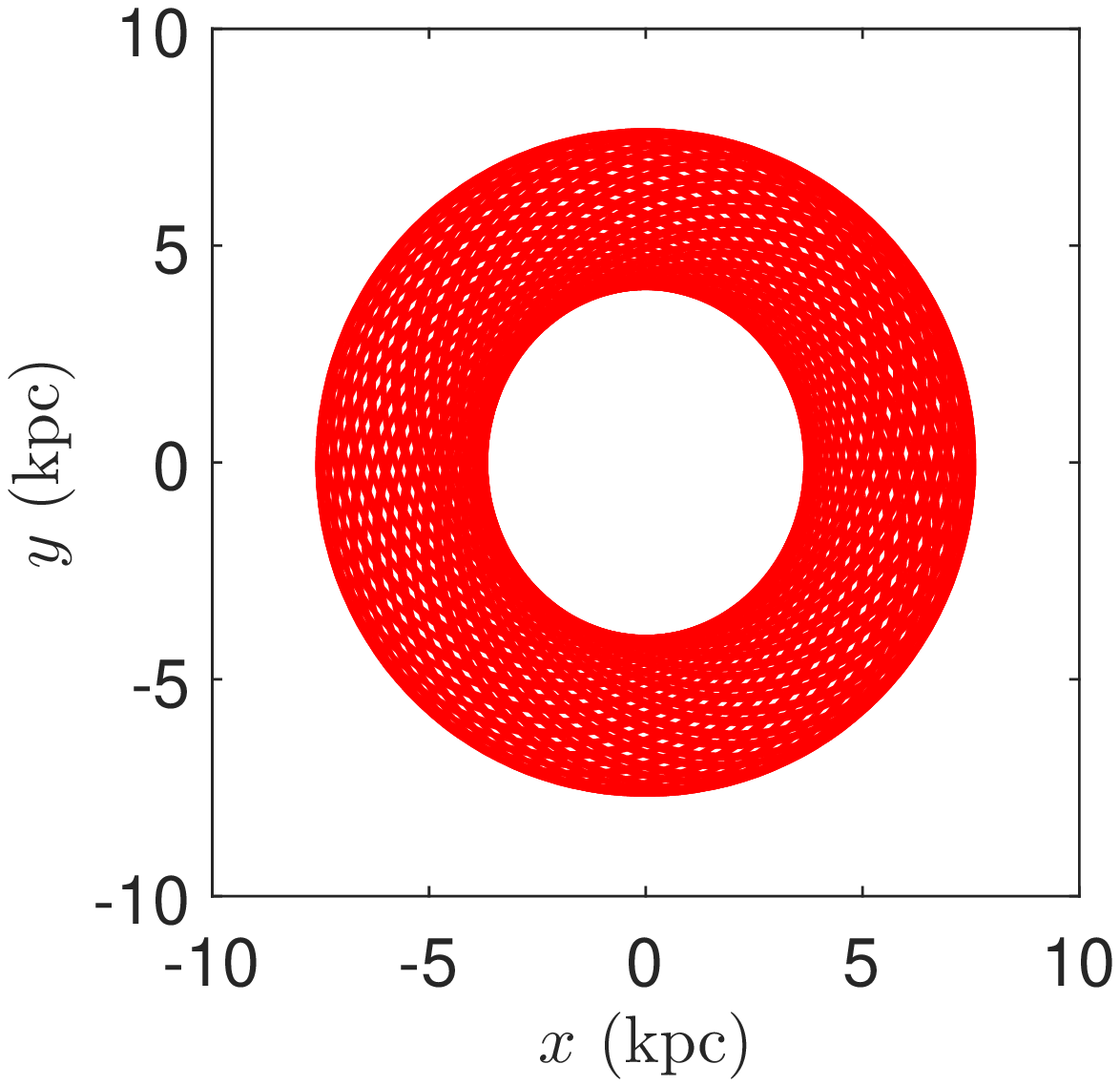}}
\caption{Model $2$: Orbits in the $x - y$ plane.}
\label{fig:3}
\end{figure}

\noindent Orbital dynamics in the $x - y$ plane are visualized for two chosen initial conditions: $(x_0,y_0,p_{x_0}) \equiv (5,0,15)$ and $(x_0,y_0,p_{x_0}) \equiv (-5,0,15)$. Now, $(x_0,y_0,p_{x_0}) \equiv (5,0,15)$ is chosen to study the orbital and escape dynamics near $L_1$ (or $L_1^{'}$), whereas $(x_0,y_0,p_{x_0}) \equiv (-5,0,15)$ is chosen to study the same dynamics at a suitable distance from $L_1$ (or $L_1^{'}$) \citep{Mondal2021}. For each of them $p_{y_0}$ value is evaluated from Eq. (\ref{eq:2}).
\begin{itemize}
\item Model $1$: For the initial conditions $x_0 = 5$, $y_0 = 0$ and $p_{x_0} = 15$ and $x_0 = -5$, $y_0 = 0$ and $p_{x_0} = 15$ stellar orbits in the $x - y$ plane are escaping chaotic and non-escaping retrograde quasi-periodic rosette orbits in nature respectively, irrespective of escape energy values (Figs. 5, 6, 7, 8 of \citet{Mondal2021}). So, for suitable choices of initial conditions near $L_1$, orbits do escape from the disc.

\item Model $2$: Figs. \ref{fig:3a} and \ref{fig:3b} show stellar orbits in the x-y plane for escape energy values of $C = 0.01$ and $C = 0.1$, respectively, for initial condition $x_0 = 5$, $y_0 = 0$ and $p_{x_0} = 15$ for model $2$. In both figures, orbits are non-escaping chaotic in nature. Similarly, stellar orbits are plotted in Figs. \ref{fig:3c} and \ref{fig:3d} for escape energy values of $C = 0.01$ and $C = 0.1$, respectively, for initial condition $x_0 = -5$, $y_0 = 0$ and $p_{x_0} = 15$. In both figures, orbits are retrograde quasi-periodic rosette in nature. $p_{y_0}$ value of each plot is evaluated from Eq. (\ref{eq:2}) and the corresponding MLE values are listed in Table \ref{tab:3}. So, for the NFW dark halo orbits do not escape from the disc even for suitable choices of initial conditions chosen near $L_1$.

\begin{table}[H]
\centering
\begin{tabular}{|c|c|c|}
\hline
Initial Condition   &$C$    &MLE\\
\hline
\hline
$(x_0,y_0,p_{x_0})$ &$0.01$ &$0.17500746$\\
$\equiv (5,0,15)$   &$0.1$  &$0.16860462$\\
\hline
$(x_0,y_0,p_{x_0})$ &$0.01$ &$0.08152241$\\
$\equiv (-5,0,15)$  &$0.1$  &$0.09217099$\\
\hline
\end{tabular}
\caption{Model $2$: MLE for different values of $C$.}
\label{tab:3}
\end{table}
\end{itemize}

\subsection{Poincaré maps}
\label{sec:3.2} 
For Poincaré surface section maps in the $x - y$ plane, we set-up a $43\times43$ grid of initial conditions with step sizes $\Delta x = 1$ kpc and $\Delta y = 1$ kpc. Among them, only those initial conditions that are within the Lagrange radius ($r_{L_1}$ or $r_{L_1^{'}}$) are considered. Also, values of $p_x$ and $p_y$ are chosen as $p_{x_0} = 0$ and $p_{y_0} > 0$ respectively, where $p_{y_0}$ is evaluated from Eq. (\ref{eq:2}). In these Poincaré surface section maps, our chosen surface cross sections are $p_x = 0$ and $p_y \le 0$ \citep{Mondal2021}. Similarly, for Poincaré surface section maps in the $x - p_x$ plane, we set-up a $43\times31$ grid of initial conditions with step sizes $\Delta x = 1$ kpc and $\Delta p_x = 10$ km $\text{s}^{-1}$. Among them, only those initial conditions that are within the Lagrange radius as already mentioned. Also, values of $y$ and $p_y$ are chosen as $y_0 = 0$ and $p_{y_0} > 0$ respectively, where $p_{y_0}$ is evaluated from Eq. (\ref{eq:2}). In these Poincaré surface section maps, our chosen surface cross sections are $y = 0$ and $p_y \le 0$ \citep{Mondal2021}.

\begin{figure}[H]
\centering
\subfigure[Poincaré surface sections of $p_x = 0$ and $p_y \leq 0$ for $C = 0.01$]{\label{fig:4a}\includegraphics[width=0.49\columnwidth]{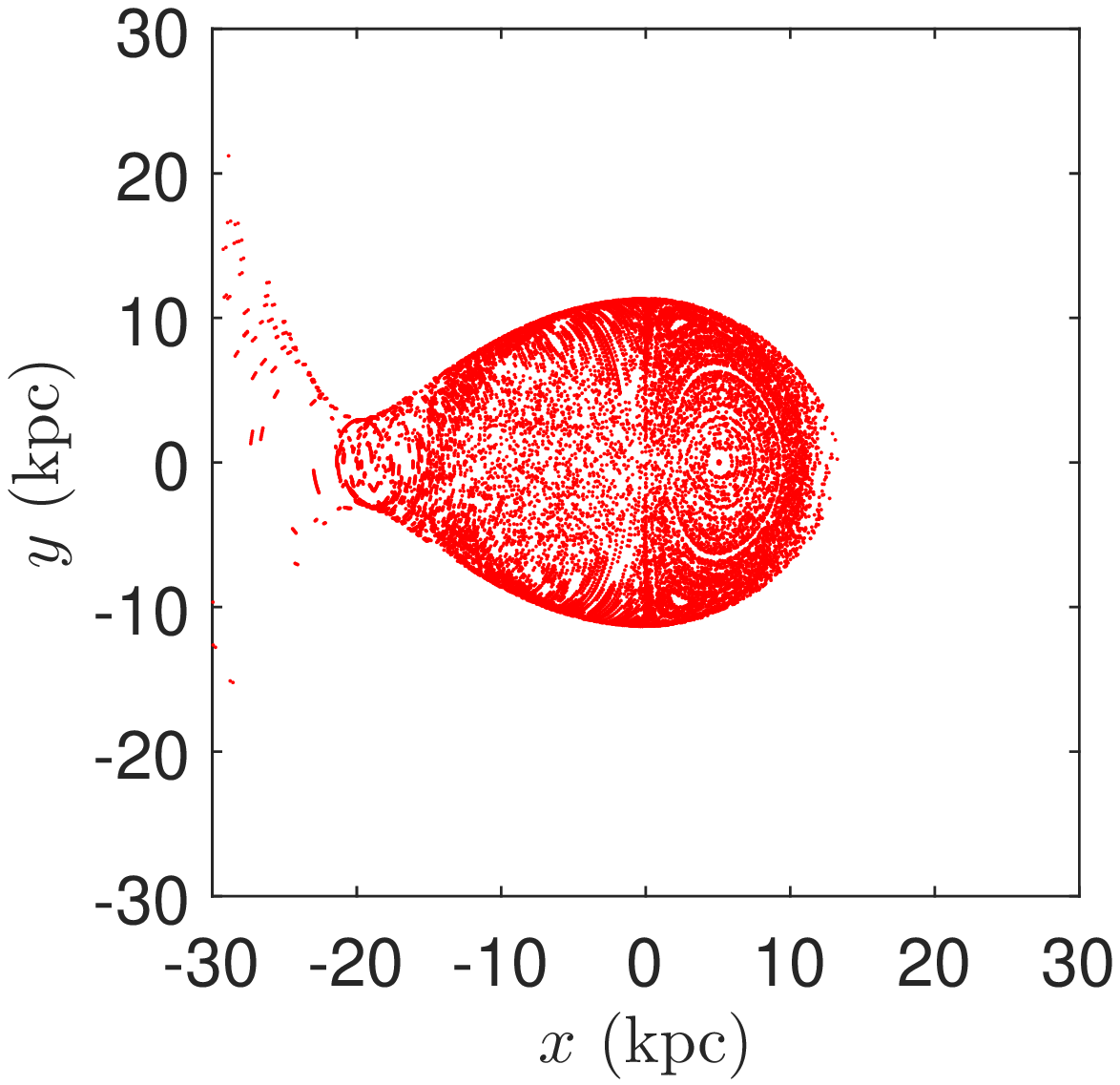}}
\subfigure[Poincaré surface sections of $p_x = 0$ and $p_y \leq 0$ for $C = 0.1$]{\label{fig:4b}\includegraphics[width=0.49\columnwidth]{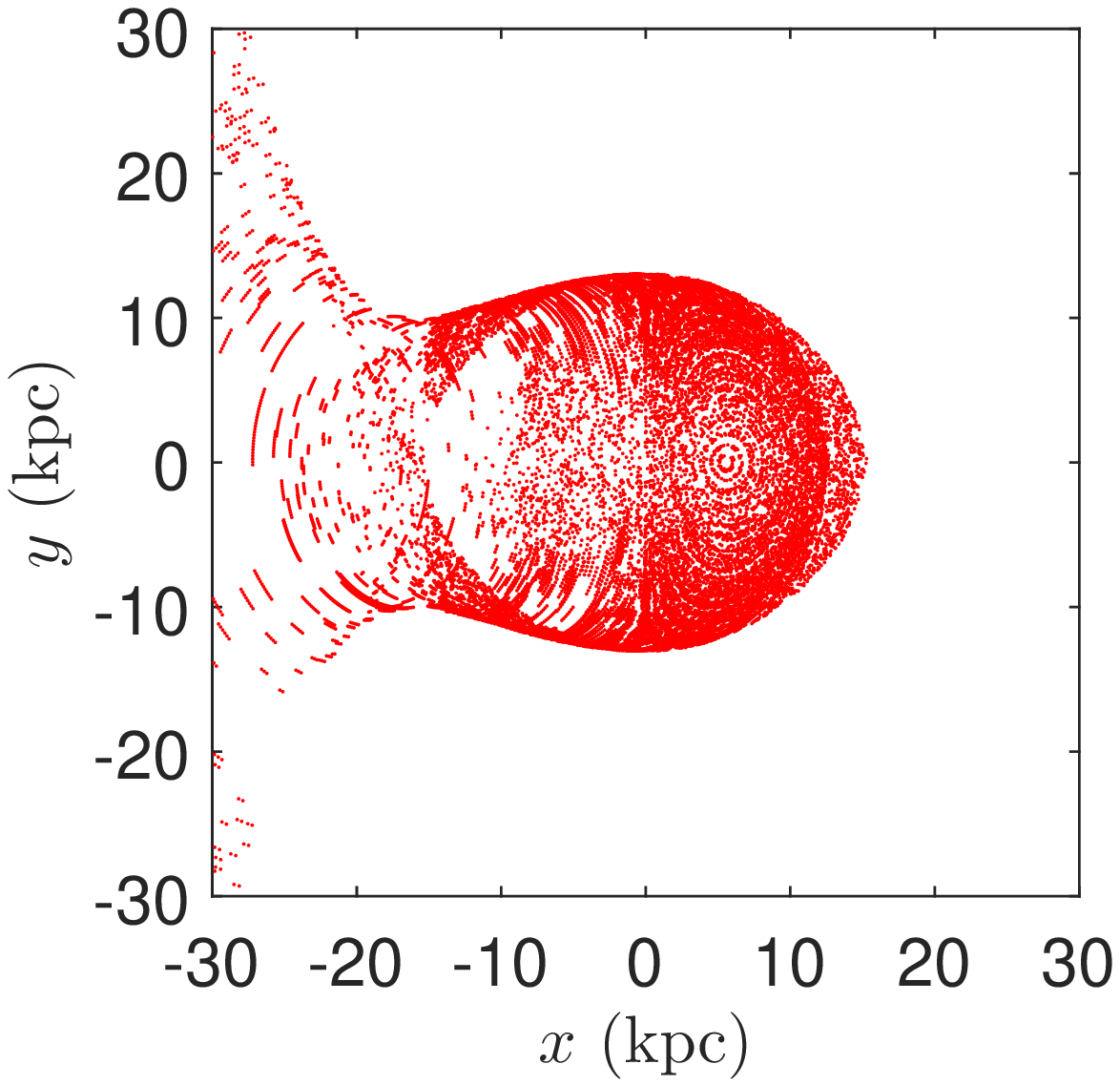}}
\subfigure[Poincaré surface sections of $y = 0$ and $p_y \leq 0$ for $C = 0.01$]{\label{fig:4c}\includegraphics[width=0.49\columnwidth]{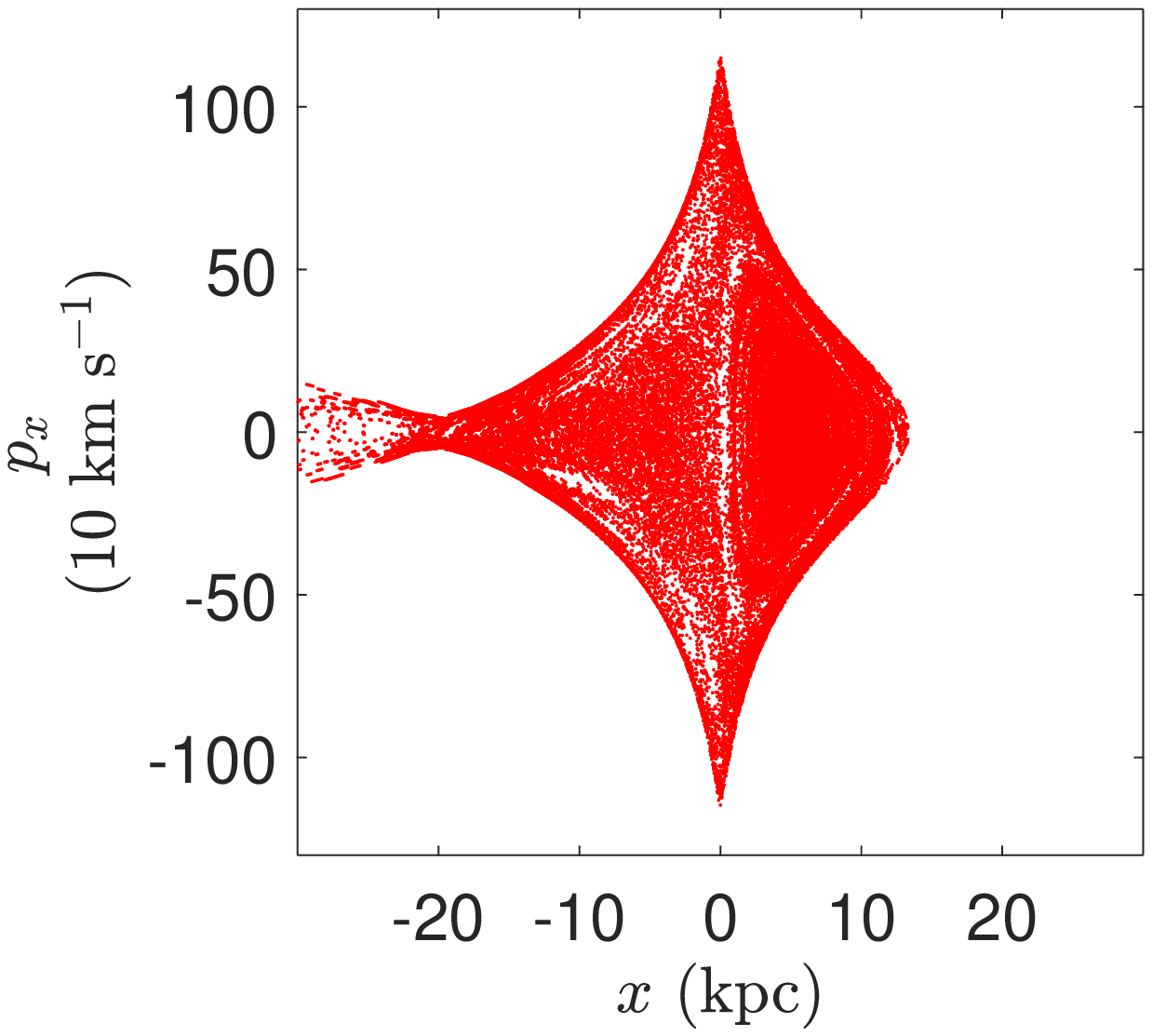}}
\subfigure[Poincaré surface sections of $y = 0$ and $p_y \leq 0$ for $C = 0.1$]{\label{fig:4d}\includegraphics[width=0.49\columnwidth]{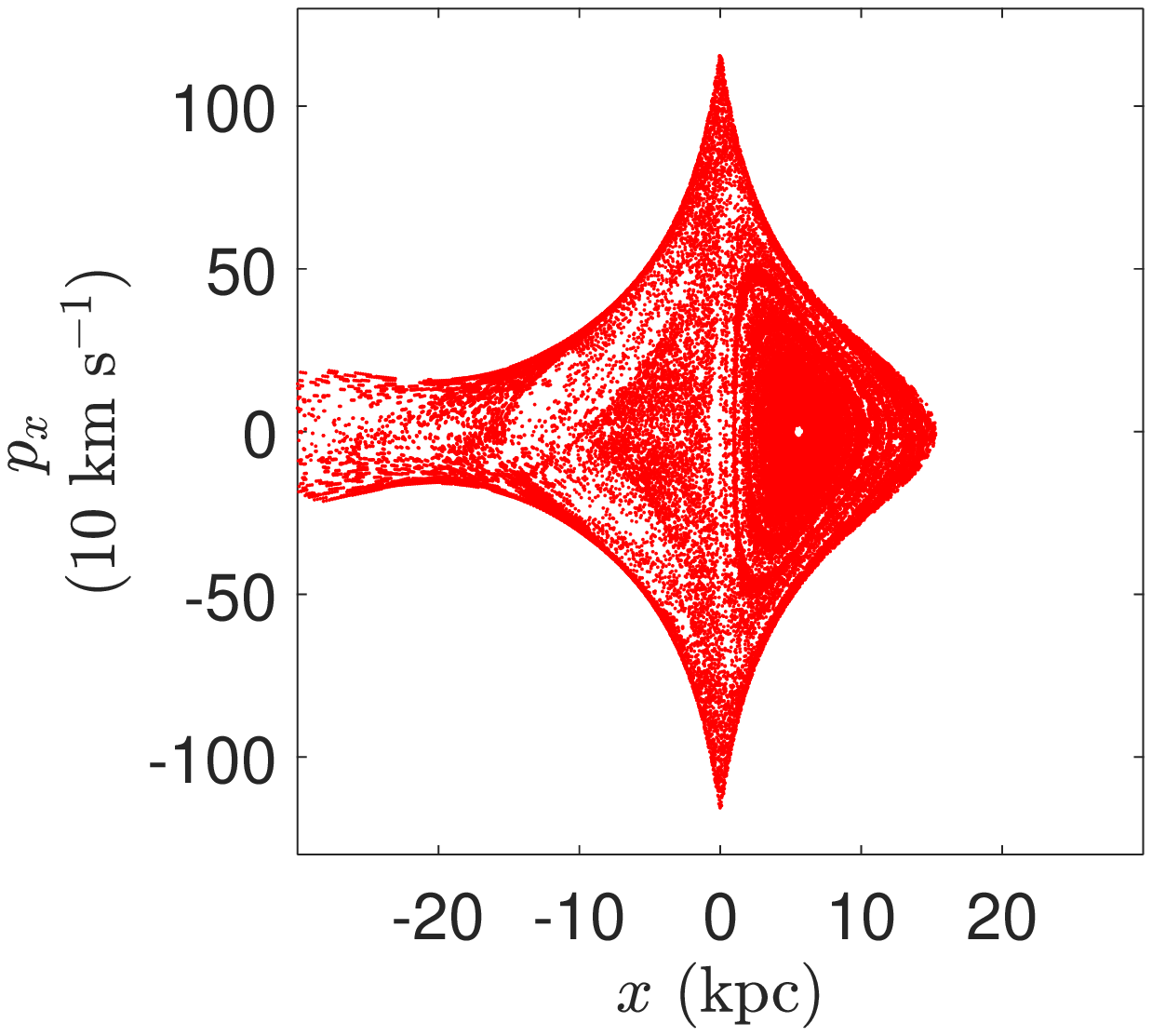}}
\caption{Model $2$: Poincaré surface section maps in the $x - y$ and $x - p_x$ planes.}
\label{fig:4}
\end{figure}

\begin{itemize}
\item Model $1$: For model $1$, Poincaré surface section maps in both the $x - y$ and $x - p_x$ planes for escape energy values of $C=0.01$ and $C=0.1$ are given in Fig. 14 of \citet{Mondal2021}. We found that a primary stability island is formed in the $x - y$ plane due to quasi-periodic motions, and the number of cross sectional points outside the corotation region increases with an increment of $C$. Similarly, a primary stability island has been observed in the $x - p_x$ plane. So, for an oblate dark halo accompanied by a strong bar, orbits may escape through the bar ends, and the amount of escape is increased with an increment in the escape energy value.

\item Model $2$: Poincaré surface section maps in the $x - y$ plane are plotted in Figs. \ref{fig:4a} and \ref{fig:4b} for escape energy values of $C = 0.01$ and $C = 0.1$, respectively, for model $2$. In both figures, we observed that a primary stability island of stellar orbits is formed near $(7,0)$ due to quasi-periodic motions. Furthermore, the number of cross sectional points outside the corotation region is increased with increment of $C$, and these numbers are large when compared to model $1$. Similarly, for escape energy values of $C = 0.01$ and $C = 0.1$, Poincaré surface section maps in the $x - p_x$ plane are plotted in Figs. \ref{fig:4c} and \ref{fig:4d}, respectively. Here also, we observed a primary stability island of stellar orbits being formed near $(7,0)$. In these figures, overall orbital trends are similar to those observed in the $x - y$ plane. The number of orbits crossing the barred region is massive as compared to model $1$. So, for the NFW dark halo accompanied by a strong bar, though the massive amount of orbits have been escaped through the bar ends, but remain confined within the disc till the integration time.
\end{itemize}

\subsection{Evolution of chaos with respect to the dark halo parameters}
\label{sec:3.3} 
Here we demonstrate the variation of mass, size, and circular velocity of the dark halo with chaos (in terms of MLE). These results have immense interest from the viewpoint of chaotic variations in the vicinity of $L_1$ (or $L_1^{'}$) with respect to the dark halo parameters. That's why, we focused only on the initial condition $(x_0,y_0,p_{x_0})$ $\equiv (5,0,15)$, where $p_{y_0}$ value is evaluated from Eq. (\ref{eq:2}).

\begin{itemize}
\item Model $1$: Fig. \ref{fig:5a} shows a variation of MLE with the dark halo flattening parameter ($\beta$) for escape energy values $C = 0.01$ and $C = 0.1$ for model $1$. Again, Fig. \ref{fig:5b} shows the variation of MLE with the circular velocity of the dark halo ($v_0$) for escape energy values $C = 0.01$ and $C = 0.1$.

\begin{figure}[H]
\centering
\subfigure[MLE for different values of $\beta$ and $C$ with $(x_0,y_0,p_{x_0})$ $\equiv (5,0,15)$]{\label{fig:5a}\includegraphics[width=0.49\columnwidth]{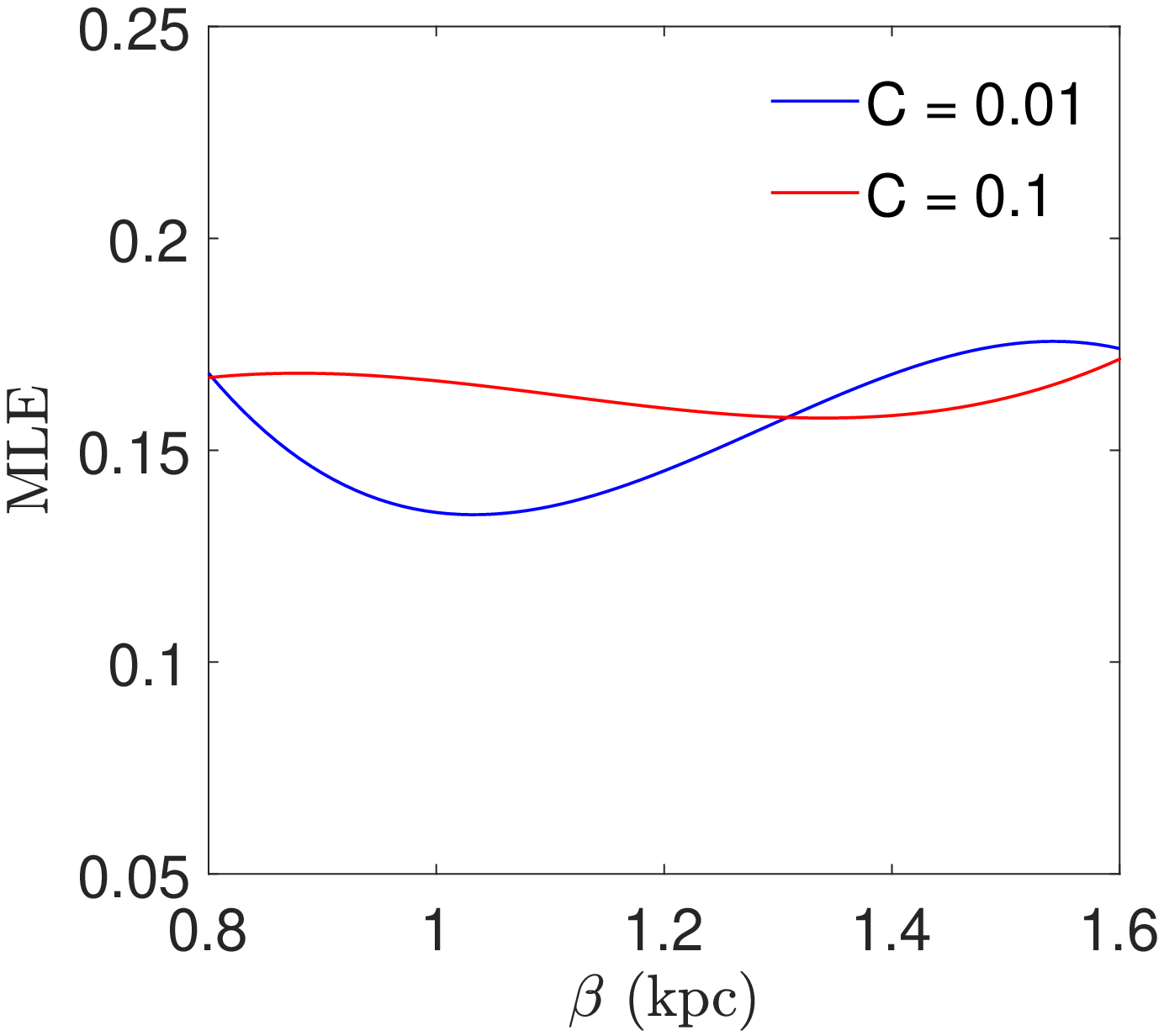}}
\subfigure[MLE for different values of $v_0$ and $C$ with $(x_0,y_0,p_{x_0})$ $\equiv (5,0,15)$]{\label{fig:5b}\includegraphics[width=0.49\columnwidth]{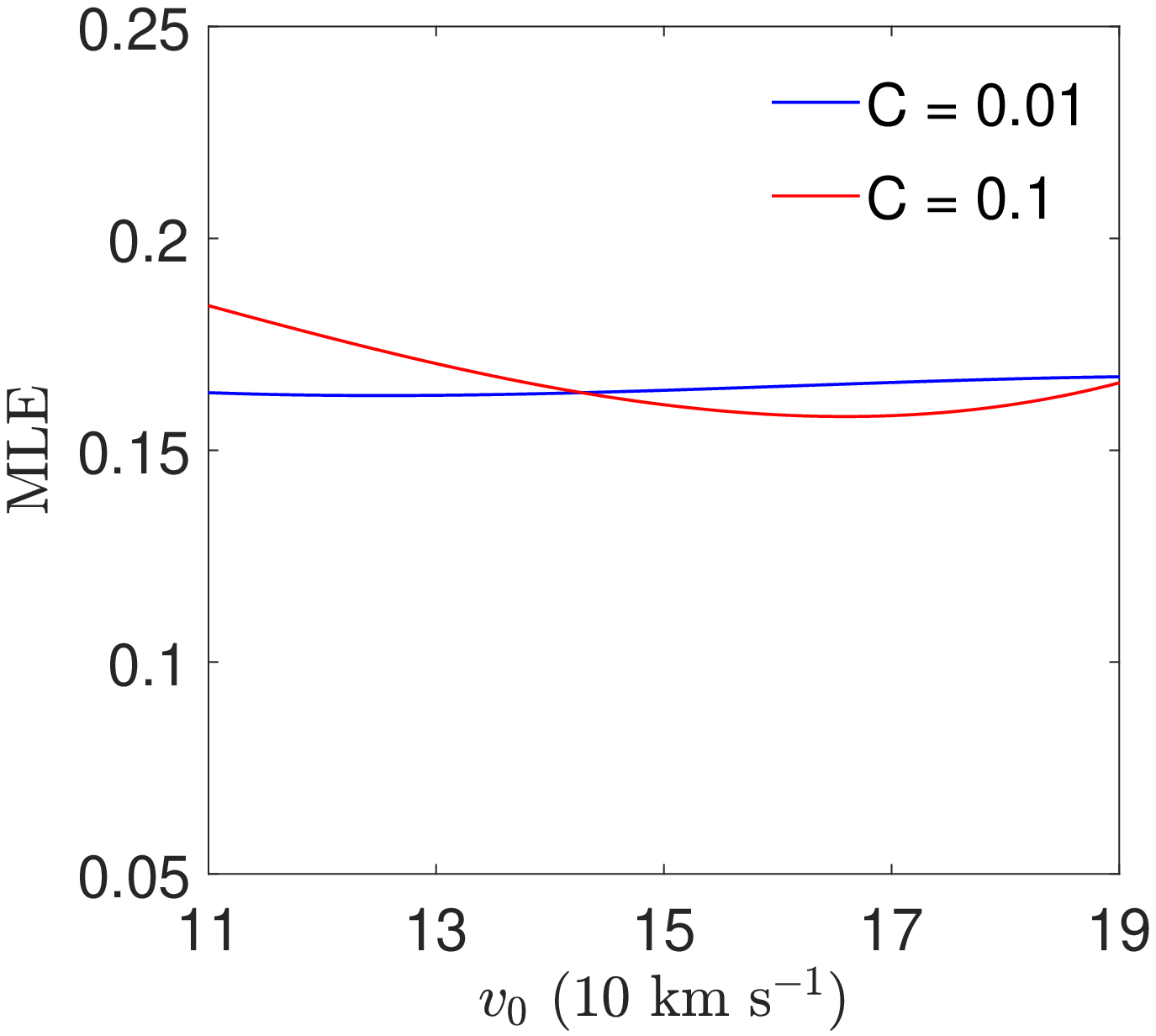}}
\caption{Model $1$: Variation of MLE with the dark halo parameters.}
\label{fig:5}
\end{figure}

\item Model $2$: For model $2$, variation of MLE with the dark halo concentration parameter ($c$) for escape energy values $C = 0.01$ and $C=0.1$ is shown in Fig. \ref{fig:6a}.Again, Fig. \ref{fig:6b} shows the variation of MLE with the virial mass of the dark halo ($M_\text{vir}$) for escape energy values of $C = 0.01 $ and $C = 0.1 $.

\begin{figure}[H]
\centering
\subfigure[MLE for different values of $c$ and $C$ with $(x_0,y_0,p_{x_0})$ $\equiv (5,0,15)$]{\label{fig:6a}\includegraphics[width=0.49\columnwidth]{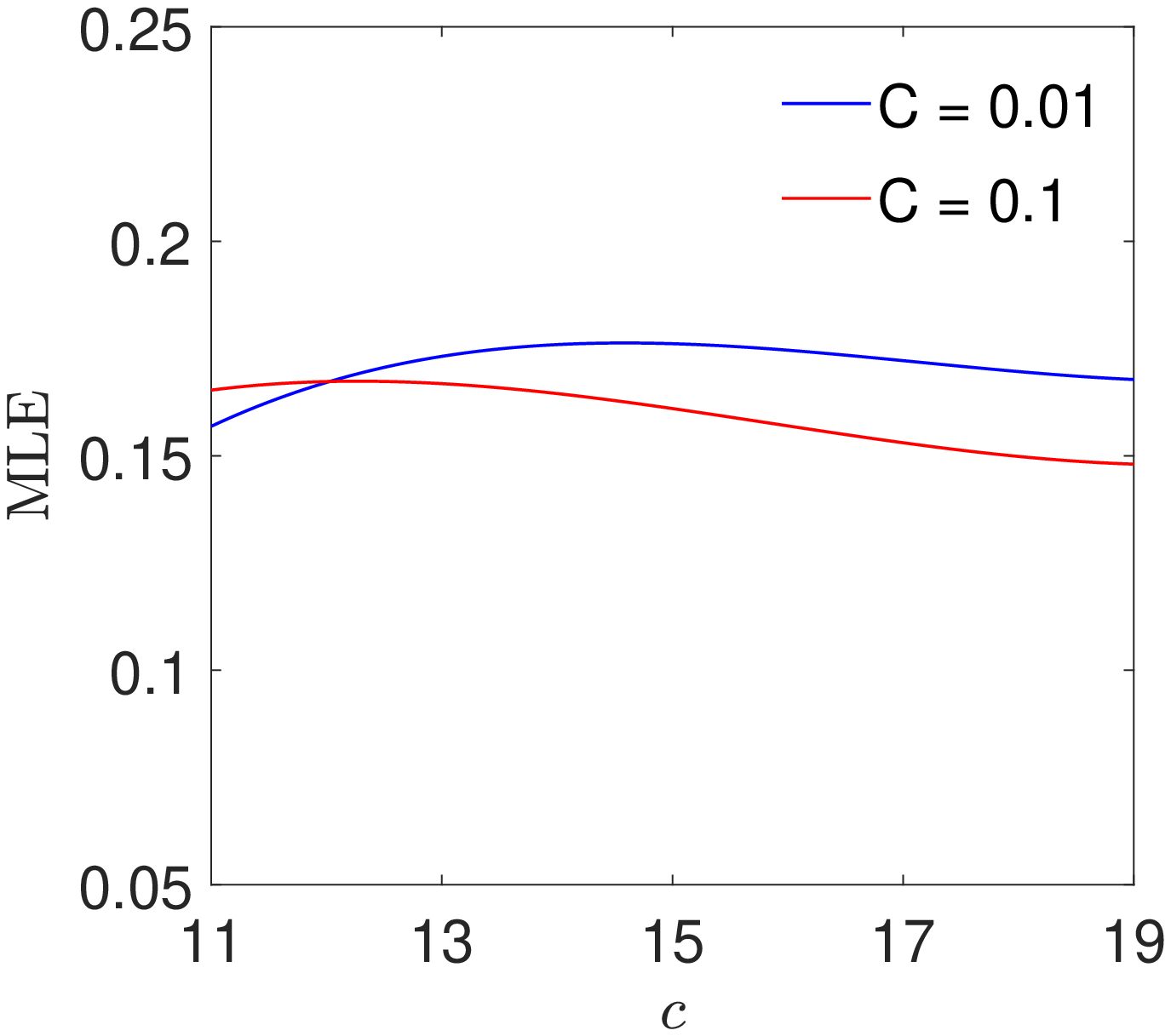}}
\subfigure[MLE for different values of $M_\text{vir}$ and $C$ with $(x_0,y_0,p_{x_0})$ $\equiv (5,0,15)$]{\label{fig:6b}\includegraphics[width=0.49\columnwidth]{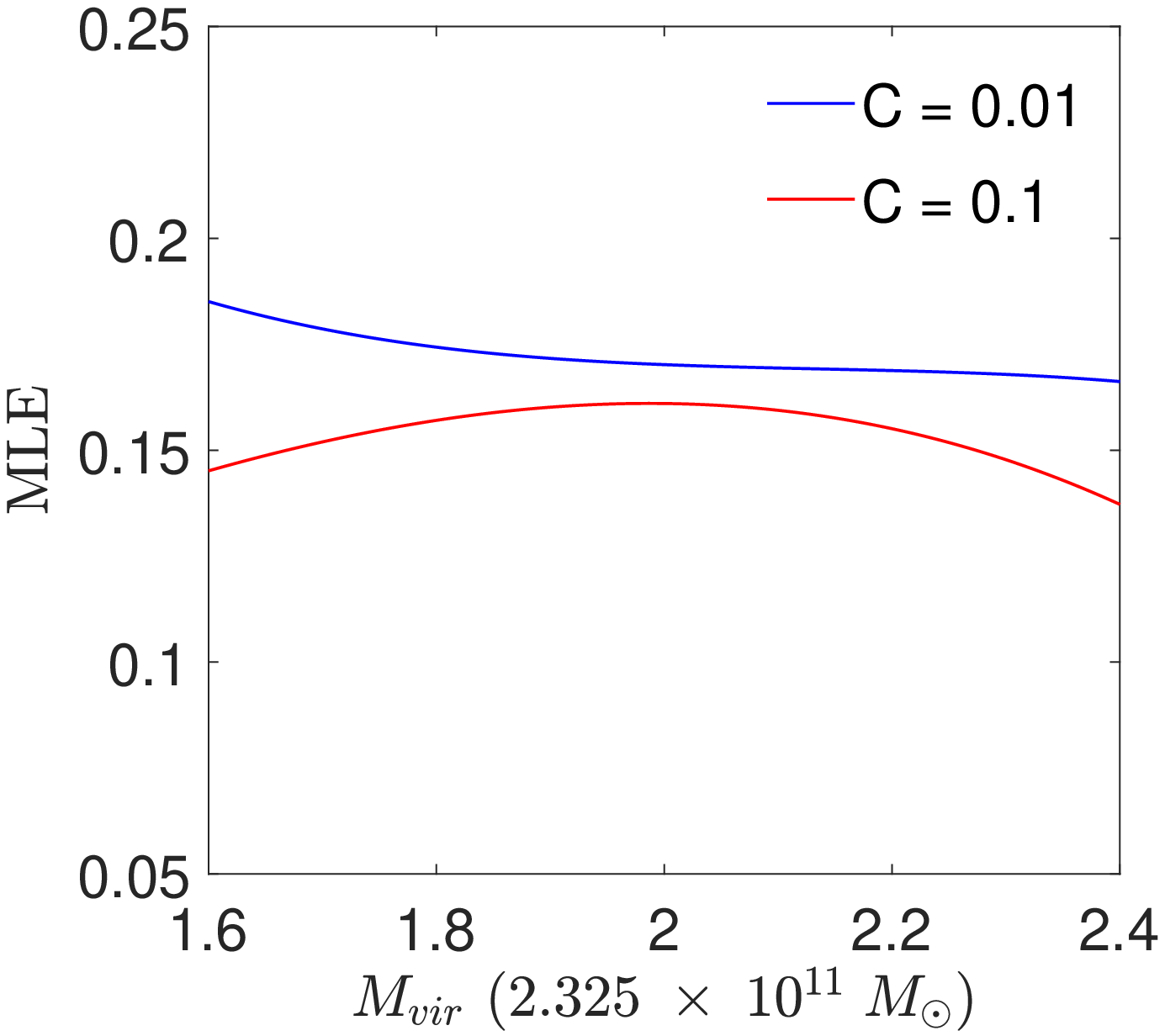}}
\caption{Model $2$: Variation of MLE with the dark halo parameters.}
\label{fig:6}
\end{figure}
\end{itemize}

\section{Discussion and conclusions}
\label{sec:4}
This work delineates the orbital and escape dynamics of stars in galaxies with strong (or massive) bars and also investigates the influence of dark halos along with the fate of escaping stars. 

\indent Our four-component gravitational model is composed of bulge, bar, disc, and dark halo. We analyzed the model for two separate dark halo profiles, namely, (i) an oblate halo and (ii) the NFW halo. The dark halo of model $1$ resembles a flat dark halo distribution (see Fig. \ref{fig:2a}), while the dark halo of model $2$ resembles a cuspy dark halo distribution (see Fig. \ref{fig:2a}). Again, in the rotation curves of both models, we observed that for model $2$ within the barred region, orbital speeds of stars and gas are higher than model $1$. While outside the barred region, that trend is reversed (see Fig. \ref{fig:2b}). This means the central cusp enhanced the orbital chaos, and that's why rotational velocities are increased within the barred region for model $2$ as compared to model $1$.

\begin{figure}[H]
\centering
\includegraphics[width=0.5\columnwidth]{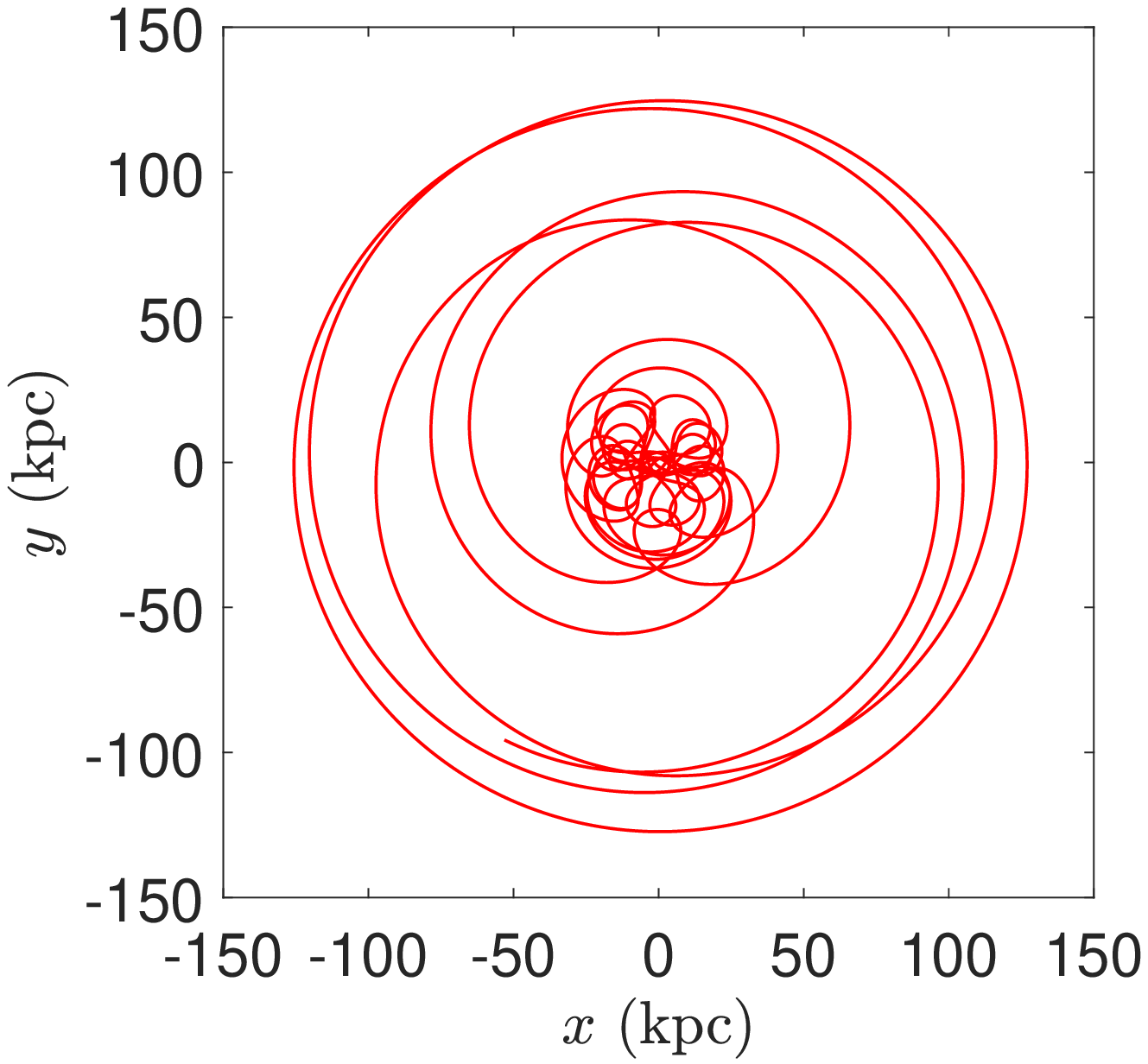}
\caption{Model $2$—escaping chaotic orbit for $C = 0.182$ with $(x_0,y_0,p_{x_0},p_{y_0})$ $\equiv (5,0,15,57.92070676)$.}
\label{fig:7}
\end{figure}

\indent Chaos (or instabilities) inside the barred region may propagate over a long time period and further relates to the formation of spiral arms or inner disc rings depending upon the bar strength (strong or weak) \citep{Contopoulos2012, Jung2016, Mestre2020, Mondal2021}. Now for model $1$, orbits are escaping and chaotic with high MLE for initial point $x_0 = 5$, $y_0 = 0$, $p_{x_0} = 15$. Again, orbits are non-escaping and quasi-periodic with low MLE for $x_0 = -5$, $y_0 = 0$, $p_{x_0} = 15$. In both cases, trends remain the same irrespective of the escape energy values (see Table 5 of \citet{Mondal2021}). So, for the oblate dark halo, stellar orbits may escape from the disc through the bar ends for suitable choices of initial starting points. On the other hand, for model $2$, orbits are non-escaping and chaotic with high MLE for initial point $x_0 = 5$, $y_0 = 0$, $p_{x_0} = 15$. Again, for $x_0 = -5$, $y_0 = 0$, $p_{x_0} = 15$ orbits are non-escaping and quasi-periodic with low MLE. Here also, in both cases, trends remain the same irrespective of the escape energy values (see Table \ref{tab:3}). So, for the NFW dark halo, stellar orbits may be confined within the disc, irrespective of initial starting points and high escape energy values. Hence, the nature of the dark halo has an immense influence on structure formation via chaotic orbits escaping from the barred region. In our analytic set-up, such chaotic orbits are found in the energy range $E > E_{L_1}$ (or $E > E_{L_1^{'}}$). Now the radial force components of both models are identical (see Fig. \ref{fig:2c}). Also, the tangential force components of both models are identical within the bulge region, but outside the bulge region tangential force component of model $1$ dominates over model $2$ (see Fig. \ref{fig:2d}). So, for the oblate dark halo, stellar orbits have a threshold tangential force, above which orbits are less escape-prone, i.e., tend to form inner disc rings rather than spiral arms \citep{Mondal2021}. Again, for the NFW dark halo, the strength of tangential force is less than the escape threshold. Hence, in model $2$, stellar orbits are more escape prone than model $1$ in order to form spiral arms (see Fig. \ref{fig:2d}). Now, the fate of escaping stellar orbits also depends upon the amount of central violence or baryonic feedback. For model $2$, though the formation of spiral arms via escape looks more promising than model $1$, but the final fate of escaping orbits needs to be further examined for more higher escape energy values. We found that orbital escape is not possible for $C = 0.01$ and $C = 0.1$ in model $2$, but possible for $C > 0.181$ (see Fig. \ref{fig:7}), which is greater than $C = 0.1$.\\	 
From all these results our observations are as follows:
\begin{itemize}
\item Model 1:\\ 
\noindent (i) From the orbital structures (Figs. 5, 6, 7, 8 of \citet{Mondal2021}), we conclude that an oblate dark halo accompanied by a strong bar encourages the orbital escapes from the disc through the bar ends (i.e., Lagrangian points $L_1$ and $L_2$). Also, Poincaré surface section maps in the $x - y$ and $x - p_x$ phase planes (Fig. 13 of \citet{Mondal2021}) show that the amount of escape has been increased with an increment in the escape energy value.

\noindent (ii) MLE values follow a stable oscillating pattern with an increase in the dark halo flattening parameter ($\beta$) for lower escape energy ($C = 0.01$). Furthermore, for higher escape energy ($C = 0.1$), MLE values vary little until $\beta < 1.3$ and then increase with an increase in $\beta$ (see Fig. \ref{fig:5a}). Giant galaxies have more flattened dark halos as these dark halos experience more expansion due to the impulsive gas outflow, i.e., escaping motion \citep{Dutton2016}. On the other hand, the amount of escape has been increased for higher values of escape energy. That's why we observe strong chaotic motions for both higher values of both $C$ and $\beta$.

\noindent (iii) Again, for lower escape energy ($C = 0.01$), MLE values very slowly increase with increment of the circular velocity of the dark halo ($v_0$). Also, for higher escape energy ($C = 0.1$), MLE values are high for $v_0 < 14$ and beyond that MLE values are low, as compared to $C = 0.01$ (see Fig. \ref{fig:5b}). We know fast rotating dark halos promote the bar's growth via angular momentum transport \citep{Saha2013}. For $C = 0.01$, the central region has a lower amount of chaos, which transports less angular momentum. As a result, we see a slow increase in chaoticity with an increment of $v_0$. But, for $C = 0.1$, more chaos in the central region transports more angular momentum, which should result in high chaoticity. In this case, only $v_0 < 14$ exhibits a high level of chaoticity when compared to $C = 0.01$. The reason for this is that for $v_0 > 14$, a large number of escapes suppresses chaos in the central region, resulting in a low amount of chaos when compared to $C = 0.01$.

\noindent (iv) The central black hole mass has a strong influence on violent activities (e.g., baryonic feedback from supernova, shocks, etc.) near the galactic center \citep{Basu1989, Efstathiou2000, Seigar2008, Berrier2013, Mondal2019}. Also, giant spirals harbor SMBHs at their centers, and as a consequence, violent activities are enhanced in their central regions. We know that the evolution of chaos for an oblate dark halo along with a strong bar well justifies the formation of full-fledged spiral arms in giant spirals if they host the central SMBH \citep{Mondal2021}. On the other hand, an enormous amount of violent activity near the galactic center may turn the central dark halo cusp into the flat core as predicted by \textit{N}-body simulations. The reason behind this is central feedback-driven gas outflows, which transfer energies to the orbits of the collision-less dark matter particles \citep{Mashchenko2006, Pontzen2012}. So, under the influence of SMBHs, central violence is amplified and the central dark halo cusp is turned into a flat core. The dark halo of model $1$ has a flat central density and, furthermore, under the influence of the central SMBH, dark halos extend beyond the visible component of the galaxy. This is the scenario of giant spiral galaxies, where full-fledged spiral arms are observed and an extended distribution of dark halos beyond the optical radius has been found. Examples of such galaxies are NGC 1300, NGC 4314, NGC 3351, etc. \citep{Sandage1961, Buta2007}. Hence, the extended dark halo distributions in the giant spirals have been well justified for the oblate dark halo-strong bar combination. Again, in the absence of the central SMBH, the oblate dark halo-strong bar combination results in less prominent or poor spiral arms \citep{Mondal2021}, where the dark halos are mostly concentrated near the center. Inability to generate sufficient feedback-driven gas outflows near the center results in such dark halo dominated cores. This is the scenario of dwarf and LSB galaxies, where less prominent or poor spiral arms are observed and dark halos are mostly concentrated in the center. Examples of such galaxies are NGC 4605, UGC 1382, etc. \citep{Simon2005, Hagen2016}. Hence, the core-dominated dark halo distributions of the dwarf and LSB galaxies have also been well justified for the same oblate dark halo-strong bar combination.

\item Model 2:\\
\noindent (i) From the orbital structures (Fig. \ref{fig:3}), we conclude that the NFW dark halo accompanied by a strong bar does not encourage the orbital escapes from the disc, but they may escape through the bar ends (i.e., Lagrangian points $L_1^{'}$ and $L_2^{'}$). Also, Poincaré surface section maps in the $x - y$ and $x - p_x$ phase planes (Fig. \ref{fig:4}) show that the amount of escape has been increased with an increment in the escape energy value. Here a massive number of orbits cross the barred region for higher escape energy values but remain encapsulated within the disc. 

\noindent (ii) MLE values increase with an increase in the escape energy (i.e., from $C = 0.01$ to $C = 0.1$) for lower values of the dark halo concentration parameter, i.e., $c < 12$. MLE values decrease for a similar increase in escape energy when $c > 12$ (see Fig. \ref{fig:6a}). The dark halo concentration parameter ($c$) relates to the dark halo density. Here, at fixed dark halo mass, $c = 12$ relates to the threshold density of the dark halo beyond which it becomes so over-dense that it does not transport angular momentum to the bar. In that instance, a further increase in the escape energy causes adiabatic gas inflow, and the dark halo contracts as a result. \citep{Dutton2016}. For this reason, chaoticity in the central region becomes suppressed for $c > 12$ at $C = 0.1$ as compared to $C = 0.01$. 

\noindent (iii) Again, MLE values decrease as the virial mass of the dark halo (($M_\text{vir}$) increases, and overall MLE values are much higher for lower escape energy ($C = 0.01$) versus higher escape energy ($C = 0.1$) (see Fig. \ref{fig:6b}). We already discussed that the dark halos of the giant galaxies experience more expansion, i.e., more escaping motion \citep{Dutton2016}. That's why excess escape results in lower chaoticity for higher values of $M_\text{vir}$. Again, the giant galaxies have more energetic centers than normal galaxies. So, in giant galaxies, more orbits escape from the barred region, which results in suppressing the chaos in the center. In normal galaxies, stellar escapes from the bar ends are comparatively much less. That's why we also observe less chaos in the central region for higher escape energies than for lower escape energies. 

\noindent (iv) We already discussed that an enormous amount of violence may eliminate the central cusp. Now, the dark halo of model $2$, i.e., the NFW dark halo has a cuspy central density. This central dark halo cusp may have evolved only into the flat core under the influence of central violence. Now, for further extension of the dark halos beyond the visible region, the system needs more extreme baryonic feedback from the central region, and that is also required to produce spiral patterns as a result of escape. We know the central regions of active galaxies have extremely energetic centers. Secondary stellar bars are also one of the possible sources of power generation in active galaxies' central engines \citep{Shlosman1989, Shlosman1990}. So the NFW dark halo profiles are better suited for galaxies with extremely energetic centers like active galaxies, and we roughly argued that double-barred systems may be one of them. 
\end{itemize}    

\indent From all the above analyses final conclusions are as follows:
\begin{enumerate}
\item The dark halo distributions in galaxies are not idiosyncratic. They evolved under the influence of central baryonic feedback, which is regulated by the central black hole mass.   

\item Stellar bars and dark halos are the two most important galactic features regarding the orbital and escape dynamics of stars. Also, the evolution of the escaping patterns substantially depends upon the natures of both stellar bars and dark halos and also on central baryonic feedback.   

\item Under the influence of strong bars, oblate dark halo profiles best describe the extended dark halo distribution and the formation of full-fledged spiral arms in the giant spirals, since they do host central SMBHs. Again, for the dwarf and LSB galaxies, the same oblate dark halo and strong bar best describe the core-dominated dark halo distribution and the formation of less prominent or poor spiral arms, since they do not host central SMBHs.

\item Under the influence of strong bars, NFW dark halo profiles do not well justify the dark halo distributions and the formation of spiral arms in most of the normal barred spirals, but they are the preferred dark halo model for the galaxies with extremely energetic centers.
\end{enumerate}

\section*{Acknowledgements}
The author DM was supported by a Senior Research Fellowship grant from the University Grants Commission of India (ID - 1263/(CSIRNETJUNE2019)). We would also like to express our gratitude to Ms. Suparna Sau (Senior Research Fellow, University of Calcutta) for several productive discussions over the MATLAB graphics in this article.
%%Also both the authors thank the anonymous referee for useful suggestions and comments which improved the work to a great extent.

\section*{Data availability}
Both authors confirm that the analyzed data supporting the our findings are included in this article and \citet{Mondal2021}.

\section*{Declarations}
Conflict of interest: The authors declare that they have no conflict of interest.



%% References


%\bibliographystyle{elsarticle-num}






% \end{thebibliography}

\section*{Appendix}
\end{document}